\documentclass[fleqn]{mnras}

\usepackage{newtxtext,newtxmath}

\usepackage[T1]{fontenc}
\usepackage{ae,aecompl}

\usepackage{graphicx}	
\usepackage{amsmath}	
\usepackage{amssymb}	

\title[How important is the diffuse continuum?]{Quantifying the impact of variable BLR diffuse continuum contributions on measured continuum inter-band delays}
\author[K.T.\ Korista and M.R.\ Goad]{
K.T. Korista,$^{1}$\thanks{E-mail: kirk.korista@wmich.edu}
M.R. Goad,$^{2}$
\\
$^{1}$Western Michigan University, Department of Physics, 1120 Everett Tower, Kalamazoo, MI 49008-5252, USA\\
$^{2}$University of Leicester, Department of Physics and Astronomy, University Road, Leicester, LE1 7RH, UK\\
}

\date{Accepted 2019 August 19. Received 2019 August 16; in original form 2019 July 2}

\pubyear{2019}

\begin{document}
\label{firstpage}
\pagerange{\pageref{firstpage}--\pageref{lastpage}}
\maketitle

\begin{abstract}
We investigate the contribution of reprocessed continuum emission (1000\AA\/ -- 10,000\AA\/) originating in broad line region (BLR) gas, the diffuse continuum (DC), to the wavelength-dependent continuum delays measured in AGN disk reverberation mapping experiments. Assuming a spherical BLR geometry, we adopt a Local Optimally-emitting Cloud (LOC) model for the BLR that approximately reproduces the broad emission-line strengths of the strongest UV lines (Ly$\alpha$ and C{\sc iv}) in NGC~5548. Within this LOC framework, we explore how assumptions about the gas hydrogen density and column density distributions influence flux and delay spectra of the DC. We find that: (i) models which match well measured emission line luminosities and time delays also produce a significant DC component, (ii) increased $\rm{n_H}$ and/or $\rm{N_H}$, particularly at smaller BLR radii, result in larger DC luminosities and reduced DC delays, (iii) in a given continuum band the relative importance of the DC component to the measured inter-band delays is proportional (though not 1:1) to its fractional contribution to the total light in that band, (iv) the measured DC delays and DC variability amplitude depends also on the variability amplitude and characteristic variability timescale of the driving continuum, (v) the DC radial surface emissivity distributions F(r) approximate power-laws in radius with indices close to $-2$ ($\approx$1:1 response to variations in the driving continuum flux), thus their physics is relatively simple and less sensitive to the unknown geometry and uncertainties in radiative transfer. Finally, we provide a simple recipe for estimating the DC contribution in disk reverberation mapping experiments.


\end{abstract}




\section{Introduction}

Multi-wavelength variability studies have proven a powerful probe of the central regions of Active Galactic Nuclei (hereafter AGN), revealing a picture of the dominant physical structures within their center and the relationships between them. Broadly speaking, these regions are from the center outwards, the accretion disk, the broad emission-line region (BLR), and the dusty torus. Each region may be associated with the dominant cooling mechanism pertinent to that location within the gravitational potential well of the super-massive black hole: thermalised continuum cooling for the disk, emission-line cooling for the BLR, and cooling by grains within the dusty torus.

Revealing the spatial distribution and kinematics of the gas and its general flow into and out of nucleus, has been a long sought goal of continuum and emission-line variability studies. In particular, it has long been known that correlated continuum and broad emission-line variability studies (reverberation mapping) can be used to determine the geometry and kinematics of the broad emission-line region, and if the emission-line gas is virialised, the mass of the central super-massive black hole. More recently, attention has focused on correlated inter-band continuum variations, as a means of mapping the disk radial temperature profile, and (if the black hole mass is known), the mass accretion rate through the disk.

In the standard disk reprocessing scenario, a geometrically-thin optically-thick disk is irradiated  from above by a compact variable X-ray source (the X-ray corona) located a few scale heights above the disk at its center (the lamp-post model). X-ray reprocessing in the disk drives correlated variations in the disk emission which propagate radially outwards (coordinated by the light travel time) exciting variations first in the hotter inner disk and then later in the cooler and larger outer disk. These variations manifest as correlated continuum inter-band delays, with longer wavelength variations lagging behind the shorter wavelength variations, and appearing both smaller in amplitude and smeared in time relative to those at shorter wavelengths. Formally, for a standard $\alpha$-disk, the disk radial temperature profile follows $T(r) \propto r^{-3/4}$, and the wavelength-dependent continuum delays $\tau(\lambda)$ follow the relation:

\begin{equation}
\tau(\lambda) \propto (M\dot{M})^{1/3}\lambda^{4/3}\; ,
\end{equation}

\noindent where $M$ is the black hole mass, and $\dot{M}$ the mass accretion rate (see e.g., Collier 2001; Cackett et~al.\ 2007). Thus, if the continuum delays can be measured with sufficient precision, and the mass of the central black hole is known, then disk reverberation importantly yields the mass accretion rate through the disk. 

Advances in data quality and program design (both in terms of S/N, temporal sampling and duration) of multi-platform, multi-wavelength reverberation mapping campaigns have now unambiguously confirmed the presence of wavelength-dependent continuum delays, the long sought after signal of continuum reprocessing in the disk, in a handful of AGN. While the data collected thus far reveal the general trend of increased delays with increasing wavelength predicted by the simple disk reprocessing scenario, in all sources, the measured delays are larger (by a factor $\approx$ few) than predicted, and exhibit enhanced delays over and above this general trend, throughout the Balmer continuum (Edelson et~al.\ 2015, 2019; Fausnaugh et~al.\ 2016; Starkey et~al.\ 2017; Cackett et~al.\ 2018; McHardy et~al.\ 2018). Two sources, notably NGC~5548, and NGC~4593, also show a sharp drop in delay immediately long-ward of the Balmer jump.  Larger than expected delays imply disk sizes which are a factor $\approx$ few too large for their luminosity. Quasar microlensing studies also reveal larger than expected disk-sizes (Morgan et~al.\ 2010; Mosquera et~al.\ 2013; Poindexter et~al.\ 2008). Together, these results raise the intriguing possibility that the standard model for accretion may in fact be a poor description of the accretion process in AGN. 

As well as the larger than expected delays, two notable observations appear particularly problematic for the ``lamp-post'' model. First, in NGC~5548 the large amplitude X-ray variations appear not to be the main driver of variations in the UV and optical continuum bands  (Gardener \& Done 2017), which are more slowly varying and reduced in amplitude than expected. Secondly, in NGC~4151, an extrapolation of the UV-optical continuum delays down to X-ray energies, indicate an abrupt disconnect; the X-ray--UV delays are far larger than the expected light-crossing time between these two regions (Edelson et~al.\ 2017). Explanations put forward to explain these results fall into two broad categories. The first aims to reduce the disk luminosity by invoking, e.g., patchy, or inhomogeneous disks (Dexter \& Agol 2011; discussed in Starkey et~al.\ 2017). The other aims to increase the delays, via e.g., scattering in a disk atmosphere (Hall et~al.\ 2018; Narayan 1996), or via secondary reprocessing regions, e.g., an EUV torus (Gardner \& Done 2017), or via contamination of the delay signal by additional variable components, e.g., diffuse continuum emission from the BLR (Korista \& Goad 2001, hereafter KG01; Lawther et~al.\ 2018). Each has merit, but no single mechanism can explain all of the phenomena we observe.


The main difficulty with disk reverberation studies heretofore is that an accurate census of the major contributors to the delay signature is sadly lacking. The accretion disk does not exist in isolation, and there are other contributing continuum sources, both variable and non-variable in flux, which contribute to the measured flux in a given continuum window. Likely most important among these is {\em continuum emission from the same gas emitting the broad emission lines} (hereafter, we will refer to as simply the ``diffuse continuum''), which should be present in AGN spectra at a significant level (e.g., KG01; Lawther et~al.\ 2018). While a simplified Balmer continuum is often modeled \textit{in isolation} in detailed AGN spectral decompositions, the longer wavelength Paschen continuum \textit{and every other physical continuum process} is to the best of our knowledge universally ignored, and effectively folded into a power law continuum fit. This makes no physical sense. 

Quantifying the diffuse continuum contribution to disk continuum inter-band delays, either by detailed modeling (KG01; Lawther et~al.\ 2018) or novel recovery techniques (Chelouche et~al.\ 2019), will be pivotal to the success of all future disk reverberation mapping studies. In what follows, we assess the DC contribution to the inter-band delays using as a point of reference the most intensively studied source, the nearby Seyfert~1.5 galaxy, NGC~5548. Uniquely, this source has been the subject of 3 major ground- and space-based reverberation mapping campaigns, and thus provides an ideal test-bed for models of the BLR and central continuum source.

In the following we extend the work of KG01 and Lawther et~al.\ (2018) to investigate quantitatively the impact of time-variable BLR diffuse continuum contributions on measured continuum inter-band delays, spanning the spectral region 1000\AA\/ -- 10,000\AA\/. In $\S$2 we present photoionisation model predictions of the diffuse continuum emitted by the broad-emission line region, using a physically-motivated model of its underlying optical-UV and ionising continuum in NGC~5548. There we explore some of the physical parameters governing the diffuse continuum's wavelength-dependent flux and delay spectra, as well as its contribution to the measured UV-optical flux and inter-band continuum delay spectra. We also explore the impact of differing characteristic timescales and amplitudes of the driving continuum flux variability on the measured inter-band continuum delays. In $\S$3 we discuss two separate considerations of geometry in its potential impact on the predictions of the flux and delay spectra of the diffuse continuum. We summarise our main findings in $\S$4, and in Appendix~A we explore additional flux contributions to the UV-optical continuum by the narrow emission line and the dusty toroidal obscuration regions.

\section{Simulations}

\subsection{Photoionisation models}
We compute several grids of photoionisation models using Cloudy, version 17.00 (Ferland et~al.\ 2017). Each two-dimensional grid consists of a range in constant total hydrogen number density, $\rm{n_H (cm^{-3})}$, slabs (``clouds'') of a particular total hydrogen column density, $\rm{N_H (cm^{-2})}$, which spans a range of incident ionising photon flux, 

\begin{equation}
\rm{\Phi_H(s^{-1}cm^{-2}, E_{ph} \geq 13.6~eV) \equiv Q_H/4\pi\/r^2}\; ,
\end{equation}
\noindent
where r is the the spherical distance of a cloud to the assumed isotropically-emitting source of ionising photons. 

In order to make more accurate predictions over the large range of gas densities in our grids, we implemented larger model atoms of hydrogen and the two ions of helium than are the defaults in Cloudy. For hydrogen all levels up through principle quantum number n = 18 were treated with fully deployed angular momentum states, while those above and through n = 48 collapsed the angular momentum states into single effective levels. For neutral and ionised helium these boundaries were n = 15 and 55 and n = 15 and 45, respectively.

We utilize the spectroscopic monitoring campaigns of the extensively studied AGN NGC~5548 ($z = 0.017175$), particularly that from the 2014 Space Telescope Optical Reverberation Mapping (STORM) campaign (De~Rosa et~al.\ 2015; Edelson et~al.\ 2015, Fausnaugh et~al.\ 2016; Goad et~al.\ 2016; Pei et~al.\ 2017; Starkey et~al.\ 2017; Mathur et~al.\ 2017) to constrain the photoionisation model predictions. For the incident underlying UV-optical continuum and ionising continuum spectrum, we adopt a spectral energy distribution (SED) approximating that of NGC~5548 as modeled in Magdziarz et~al.\ (1998). This model includes a physical Comptonised accretion disk spectrum and X-ray power law, similar to the SED presented in Mehdipour et~al. (2015), and represents a substantial improvement over the rough empirical model of Walter et~al.\ (1994) adopted by KG00 and KG01. This is particularly important for establishing the level of the underlying UV-optical continuum and so the contribution of the diffuse continuum to the measured continuum flux in this well-studied AGN. 

Here we assume a luminosity distance of NGC~5548 of 72~Mpc, and coupled with a measured mean continuum flux at $\lambda$1157\AA\/, $\rm{ f_{\lambda\/1157} = 52.41\times10^{-15} ergs\, s^{-1} cm^{-2} }$ \AA\/$^{-1}$ (corrected for Milky Way extinction), obtain the corresponding far-UV and ionising luminosities (ergs s$^{-1}$) $\rm{\log \lambda\/L_{\lambda\/1138} = 43.5866}$, $\rm{\log L_{ion} = 44.3369}$, and $\rm{\log Q_H(s^{-1}) = 54.1266}$, for the adopted SED. The hydrogen ionising photon flux incident at the illuminated face of clouds lying 12.6~light-days from the central source is then $\rm{\Phi_H = 10^{20} }$ s$^{-1}$ cm$^{-2}$. All measured fluxes are corrected for grain extinction in the Milky Way assuming $A(V) = 0.055$ for NGC 5548 (Schlafly \& Finkbeiner 2011), $R_{V} = 3.1$, and the Fitzpatrick (1999) Milky Way extinction curve (Fausnaugh et~al.\ 2016). 

Unless otherwise noted, the computed photoionisation model grids assume the abundance set adopted in Korista \& Goad (2000; hereafter KG00) for the BLR in NGGC~5548: $0.5\times$ solar metallicity, except solar values in C/H and N/H. This was done in the interest of continuity, as regards to the results presented in KG01. In $\S$2.2, we compare predictions of the diffuse continuum between that abundance set and those adopting solar abundances.

In Figure~\ref{plot_phi_nh} we map out logarithmic contours of the predicted flux ratio $ \rm{ \lambda\/ F_{\lambda} }$(diffuse cont.)/$\rm{\lambda\/F_{\lambda\/1215}}$(incident cont.) in the cloud gas density--incident ionising photon flux plane for a grid of photoionised clouds with a fixed hydrogen column density $\rm{\log N_{H}(cm^{-2})=23}$.  We emphasise that the predicted emission within all bands includes all important physical processes associated with continuous emission by BLR clouds: free-bound, free-free, and scattering of the incident continuum modulated by absorption opacities. As presented in Figure~\ref{plot_phi_nh}, each cloud's emission is for full central ionising source coverage. Lines of constant ionisation parameter $\rm{\log({U_Hc}) \equiv \log(\Phi_H/n_H)}$ lie at 45 degree angles, and decrease in value from upper left (where the clouds are fully ionised and isothermal at their Compton temperatures) to the lower right. In each panel the star is a reference point locating a cloud with total hydrogen gas density of $10^{10}$ cm$^{-3}$ and ionisation parameter of $10^{-2}$, and the triangle marks the cloud parameters for the maximum in this ratio. (We note that for the $\lambda$1461 band the solid contour representing the minimum flux ratio of 1 appears twice, with local maxima appearing near the lower left and upper right corners of the grid.) The diffuse continuum is generally dominated by H, He free-bound continua, but also includes free-free emission, electron scattering, and scattering by neutral hydrogen (``Ly$\alpha$ Rayleigh scattering''; see Korista \& Ferland 1998). The electron scattering optical depth is always $\lesssim 0.07$ for the $\rm{ N_{H}=10^{23} cm^{-2} }$ clouds presented in Figure~1. We refer the reader to Korista \& Goad (2001; hereafter KG01) for additional details concerning the physics of the continuum emitted and scattered by the BLR clouds. The primary story told in Figure~\ref{plot_phi_nh} is that the cloud continuum emission generally strengthens at higher gas densities, as the emission lines thermalise at high densities and optical depths\footnote{Low-density photoionised gases (H~II regions, PNe, the narrow emission lines regions of AGN, etc.) are \textit{weak} UV-optical continuum emitters primarily because they cool much more efficiently via bound-bound, particularly electric-dipole forbidden, transitions.}. 

\subsection{A steady-state model of the BLR}

\begin{figure}
\includegraphics[width=\columnwidth]{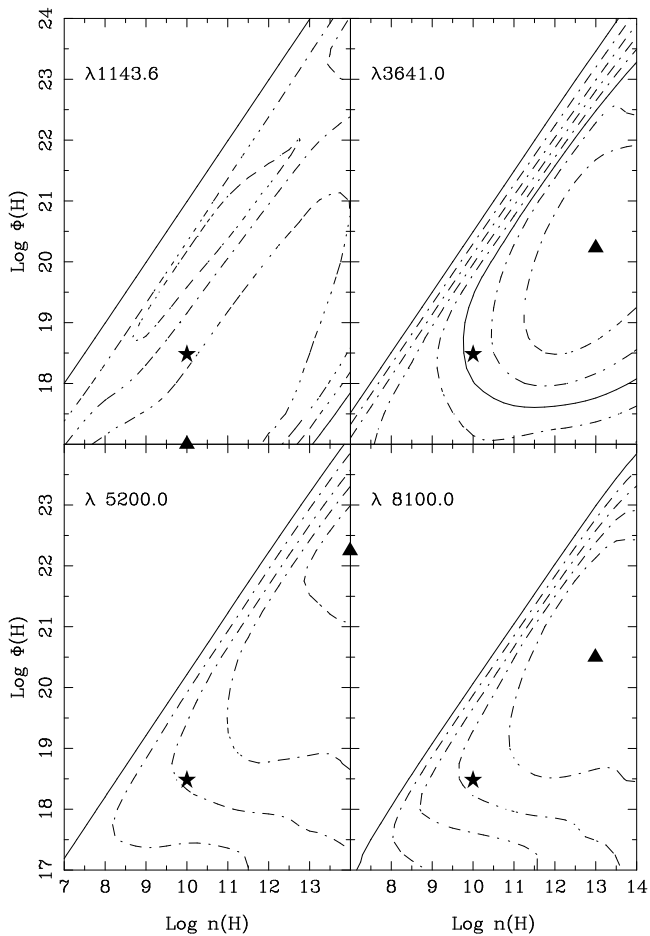}
\caption{
Logarithmic contours of the ratio $\rm{\lambda\/F_{\lambda}}$(diffuse cont.)/$\rm{\lambda\/F_{\lambda\/1215}}$(incident cont.) in the cloud gas density-incident ionising photon flux plane for a grid of photoionised clouds with a fixed hydrogen column density $\rm{\log N_{H}(cm^{-2})=23}$, at four representative wavelengths typically measured in disk and broad emission-line reverberation mapping campaigns. In each panel the minimum flux ratio contour has a value of 1 (lying near the centre-most solid diagonal), and the contours increase generally in the direction of increasing gas density (along a trajectory from lower left to the upper right) in 0.2 dex intervals (dot-dashed). See text for further details. }
\label{plot_phi_nh}
\end{figure}

Since direct observational constraints upon models for the diffuse continuum emission are few, we decided to explore the contribution and variable behavior of the BLR diffuse continuum within the simple spherical model BLR geometry for NGC~5548 presented in KG00. See also Kaspi \& Netzer (1999) for a similar model geometry. While the geometry of the BLR is unlikely to be spherically distributed about the central source of ionising photons (Horne et~al.\ 1991; Pancoast et~al.\ 2012,2014,2018; Grier et~al.\ 2012, 2013, 2017; Bentz et~al.\ 2010; Skielboe et~al. 2015), its symmetry requires the fewest adjustable parameters. This allows its predictions to be more general and not peculiar to a particular geometry or our line of sight orientation to it. And as we will show, this particular model from KG00 has both sufficient energy in its stronger broad emission lines and lags in the integrated emission line fluxes that are approximate matches to those measured by De~Rosa et~al.\ (2015) and Pei et~al.\ (2017). While we certainly do not claim this simple model to be a definitive one for NGC~5548, it is critical that any model exploring the continuum emanating from the BLR predicts the appropriate mean luminosity and variability behavior as the emission lines \textit{originating in the same gas}. The broad emission lines, are thereby able to serve as constraints to any model predictions concerning the diffuse continuum from the BLR. 

The model steady-state spherical BLR described in KG00 has an inner radius of 1~light day and an outer radius of 140~light days. In KG00, this radial distance was then thought to lie near the dust grain sublimation radius. The predicted Mg~{\sc ii} $\lambda$2798 and C~{\sc iii}] $\lambda$1909 luminosities were the most sensitive to the model's outer radius. We emphasise that the measured dust {\em delays} in NGC~5548 of 40-80 days (Suganuma et~al.\ 2006; Koshida et~al.\ 2014; Landt et~al.\ 2019) {\em do not} directly measure (1:1) physical {\em distances} $R_{dust}$ to the hot grains, particularly since the thermally-emitting grains are expected to belong to a geometry with substantial scale-height, and thus with emission from gas uplifted into the observer's direction and lying near the line of sight with substantially smaller time delays than $R_{dust}$/c. Additionally, for a given geometry, the measured cross-correlation function lags depend also upon the characteristic time scale of the continuum fluctuations. See Goad, Korista, \& Ruff 2012; Goad \& Korista 2014; Ramolla et~al.\ 2018.  The full cloud geometry covers 50\% of the sky as seen from the ionising continuum source (assuming a power-law index for the differential covering fraction relation, $dC(r)/dr \propto r^{c}$ with $c=-1.2$). To establish a point of reference, we adopt the same ``Locally Optimally-emitting Clouds'' (LOC; Baldwin et~al.\ 1995) emission line model presented in KG00 and KG01. In particular, the emission from clouds of a fixed column density $\rm{ {N_H} = 10^{23} cm^{-2} }$ (for simplicity) with a range of (constant) hydrogen number density of $10^8-10^{12} \rm{cm^{-3}}$ and spanning a range in ionisation parameter $6 \leq \log(\rm{U_Hc}) \leq 11.25$ were summed to contribute to the spectrum. We refer the reader to Bottorff et~al.\ (2002) for a generalized description of integrating over cloud parameters in the LOC picture, and to KG00 for specifics related to the adopted LOC model. Alternate viable photoionised cloud models include the constant total pressure or radiation pressure-confined clouds (RPC; Baskin, Laor, \& Stern 2014), and radially-dependent pressure law cloud models (e.g., Rees, Netzer \& Ferland 1989; Goad, O'Brien \& Gondhalekar 1993; Kaspi \& Netzer 1999, and Lawther et~al.\ 2018). Other proposed geometries for the BLR in NGC~5548 include a hydromagnetically driven disk-wind  (Bottorff et~al.\ 1997), a radiatively driven disk-wind (Chiang \& Murray 1996), a nest-shaped (Mannucci et~al.\ 1992) geometry, and that approximating the surface of a bowl with a center to rim extent of 100 light days (Goad, Korista, \& Ruff 2012; Goad \& Korista 2015).

\begin{figure}
\includegraphics[width=\columnwidth]{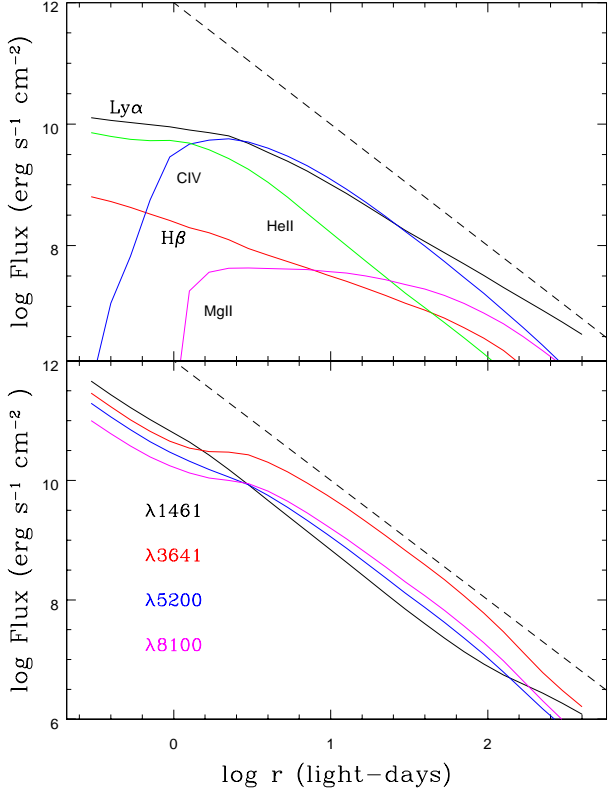}
\caption{
Upper panel -- Radial surface emissivity distributions $F(r)$ for the strongest UV and optical broad emission lines. Lower panel -- example diffuse continuum bands. The dashed line indicates a power-law emissivity function $F(r) \propto r^{\gamma}$, with logarithmic slope $\gamma = -2$, indicative of a radial responsivity distribution $\eta(r)=1.0$. $F(r)$ for Mg~{\sc ii} is relatively flat over much of the BLR, giving rise to large emissivity-weighted and responsivity-weighted radii, and consequently a low amplitude response to continuum variations.}
\label{surface_emissivities} 
\end{figure}

Example predicted radial surface emissivity distributions, $F(r)$ (erg~s$^{-1}$~cm$^{-2}$), for several of the often measured UV-optical broad emission-lines and four diffuse continuum bands, determined by summing over the model grids, are shown in Figure~\ref{surface_emissivities}. To guide the eye, we also indicate a simple power-law with slope $-2$ (dashed line). Note that for He~{\sc ii}~$\lambda$1640, $F(r)$ is a good approximation to a power-law of slope $-2$, except in the inner BLR ($r < 2$ light-days), the latter primarily a consequence of our chosen cut-off in gas hydrogen density. Extending the range in gas densities to larger values (e.g., $10^{13}$~cm$^{-3}$) extends the power-law behavior in surface emissivity to smaller $r$, and acts favorably to reduce the mean formation radius for this line, thus also increasing its response amplitude. Similarly, $F(r)$ for the majority of the diffuse continuum bands can also be approximated by a power-law in radius with slope $-2$ (Figure~\ref{surface_emissivities}, lower panel). This behavior is a manifestation of the open contours of $\rm{\lambda\/F_{\lambda}}$(diffuse cont.)/$\rm{\lambda\/F_{\lambda\/1215}}$(incident cont.) within the hydrogen ionising photon flux $\Phi_{\rm H}$, gas hydrogen density $n_{\rm H}$ plane (Figure~\ref{plot_phi_nh}). The DC bands just short-ward of the Balmer and Paschen jumps show the largest, yet still modest, deviation from this simple behavior.

In Table~1 we report the integrated broad emission-line strengths for the strongest UV and optical broad emission-lines calculated by integrating over their radial surface emissivity distributions and radial differential covering fraction dependence, for the new photoionisation computations reported here and the LOC cloud distribution model and geometry reported in KG00. The model Ly$\alpha$ luminosity\footnotetext{This model predicts additional contributions from He~{\sc ii} $\lambda$1216 and O~{\sc v}] $\lambda$1218, which sum to an additional 6.6\% in the luminosity of the broad emission line feature centered on 1216\AA\/.} reported in Table~1 finds an equivalent width of the steady state (mean) spectrum of $\approx$140\AA\/, and a Ly$\alpha$/C~{\sc iv} luminosity ratio of 1.16. Based on the spectral model fits reported in Kriss et~al.\ (2019), we find a mean measured broad Ly$\alpha$ emission line equivalent width during the AGN STORM campaign of $\approx$150\AA\/, and a broad Ly$\alpha$/C~{\sc iv} luminosity ratio of 1.13. We also report for these broad emission lines their emissivity-weighted and responsivity-weighted\footnote{The radial emission line  responsivity, $\eta(r)$, is a measure of the rate of change in flux of an emission line or diffuse continuum band relative to the flux variations in the driving ionising continuum. See Korista \& Goad (2004).} radii, indicative of a characteristic ``mean'' formation radius and response timescale (via $R \sim c<\tau>$) for each line. Note these values are broadly representative of steady-state values only, i.e., those that would be realised if the BLR is probed by continuum variations which are large enough in amplitude to excite a response in the emission line, and with a characteristic timescale which is as long as, if not longer, than the light-crossing time to the region where the line is predominantly formed (Goad \& Korista 2014).

\begin{table}
\centering
\caption{Model broad emission-line steady-state luminosities and characteristic BLR ``sizes'' for the KG00 LOC model of NGC~5548.}
\label{tab1}
\begin{tabular}{lcccrr}  \hline 

Line ID & $^{\ddagger}$log L & $R_{\rm ew}$ &  $R_{\rm rw}$  \\
        &  (erg/s) & (lt-days) & (lt-days) \\ 
\hline
Ly$\alpha$ $\lambda$1216 & 42.636 & 43.3 & 46.3 \\
Si~{\sc iv} $\lambda$1397 & 41.299 & 23.8 & 31.8 \\
O~{\sc iv}] $\lambda$1402 & 41.233 & 27.3 & 34.6 \\
C~{\sc iv} $\lambda$1549 & 42.573 & 31.8  & 40.0 \\
He~{\sc ii} $\lambda$1640 & 41.779 & 21.3 & 24.7 \\
O~{\sc iii}] $\lambda$1663 & 40.951 & 40.3 & 49.0 \\
Mg~{\sc ii} $\lambda$2798 & 41.744 & 69.1 & 83.5 \\
H$\beta$ 4861 & 41.410 & 59.5 & 69.4 \\
\hline
\end{tabular} 
\newline
%


\noindent $^{\ddagger}$ Quoted broad emission-line luminosities assume 50\% coverage.
\end{table}
%
%

\begin{figure}
\includegraphics[width=\columnwidth]{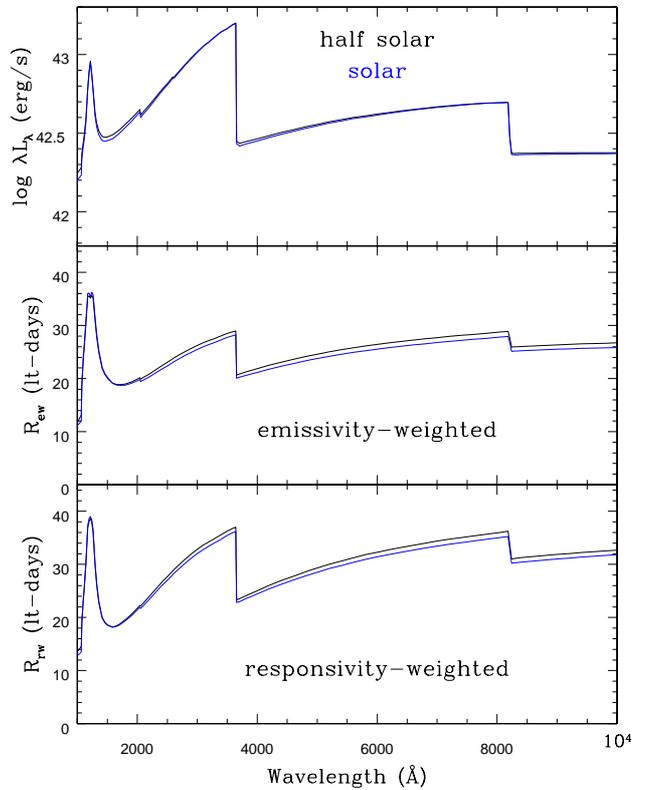}
\caption{Upper panel -- Wavelength-dependent diffuse continuum band luminosities $\log_{10} \lambda L_{\lambda}$~(erg~s$^{-1}$) for our steady-state model assuming 50\% coverage of the continuum source. Middle panel -- the predicted emissivity-weighted radii, $R_{\rm ew}$ (effectively the half-light radius), for the steady-state model. Lower-panel -- the predicted responsivity-weighted radius, $R_{\rm rw}$ (see text for details). Black and blue lines denote the elemental abundances adopted in KG00/KG01 (heavy elements half solar, except solar C/H and N/H) and solar abundances, respectively.}
\label{dc_halfsolar}
\end{figure}

\begin{figure}
\includegraphics[width=\columnwidth]{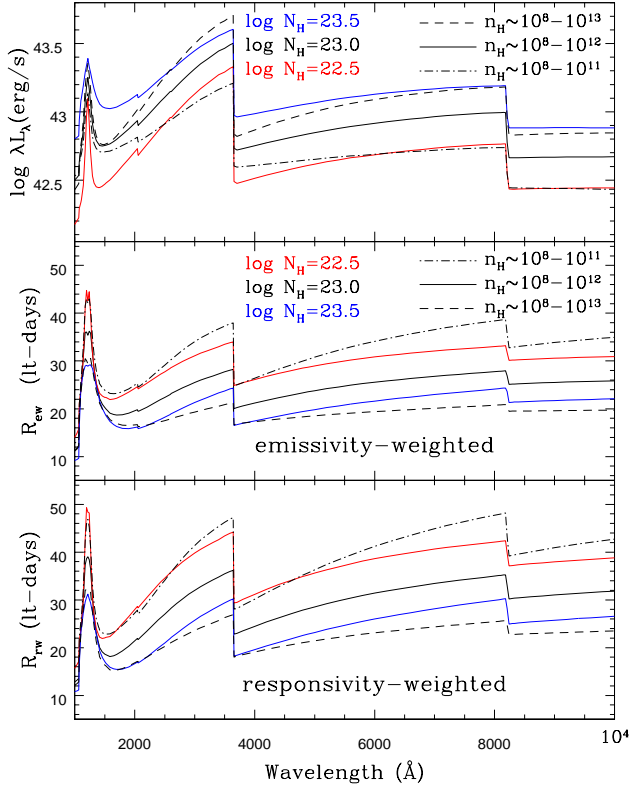}
\caption{Upper panel --  the steady-state wavelength-dependent luminosity $\lambda$L$_{\lambda}$ predicted by LOC models of three fixed values of the hydrogen column density $\log {\rm N_H (cm^{-2})}=23$ (black), 22.5 (red), 23.5 (blue), and for $\log {\rm N_H (cm^{-2})}=23$ three ranges in hydrogen gas density $\log \rm{n_H (cm^{-3})}$ (dot-dashed lines: 8--11; solid lines: 8--12; dashed black line: 8--13). Middle panel -- the corresponding wavelength-dependent emissivity-weighted radius in light days. Lower panel -- the corresponding wavelength-dependent responsivity-weighted radius in light days. See text for details.}
\label{dc_solar}
\end{figure}

For the steady-state model (centered on the mean ionising luminosity of NGC~5548, noted in $\S$2.1) we also compute the BLR's diffuse continuum spectrum, the luminosity $\rm{\lambda\/L_{\lambda}}$ as a function of wavelength for the diffuse continuum arising from BLR clouds (Figure~\ref{dc_halfsolar}, upper panel). We also calculate the emissivity-weighted and responsivity-weighted radius ($R_{\rm ew}$ and $R_{\rm{rw}}$, respectively) for each diffuse continuum band. The last of these should be representative of the predicted delay signature for an idealized driving continuum, one whose characteristic variability timescale is able to probe regions with large spatial extent. These are shown in Figure~\ref{dc_halfsolar} (middle and lower panels respectively). First, we point out that all three wavelength-dependent quantities are insensitive to the two sets of elemental abundances (described in $\S$2.1), as expected since the diffuse continuum is dominated by hydrogen. Unless the gas metallicity differs significantly from solar, we expect only minor dependencies through small differences in the electron temperatures within the clouds. Next, the DC luminosity shows significant wavelength dependence, generally increasing through the Balmer and Paschen continua and peaking at the Balmer jump. Similarly, the emissivity- and responsivity-weighted radii show a significant wavelength dependence generally increasing toward longer wavelengths, with abrupt declines long-wards of the Balmer and Paschen jumps. Note that for the KG00 model BLR the emissivity- and responsivity-weighted radii are fairly large, spanning $\approx$20--40~light-days in the UV-optical continuum bands. In all cases, the responsivity-weighted radius at a given wavelength is slightly larger than the corresponding emissivity-weighted radius (Figure~\ref{dc_halfsolar}, lower panel) because the local responsivity is generally less than 1 throughout the inner portions the BLR (Goad, O'Brien, \& Gondhalekar 1993; Korista \& Goad 2004)\footnote{If the marginal response to the driving continuum variations is constant with radius, then $R_{\rm ew}=R_{\rm rw}$ by definition.}. Because the radial surface emissivities of the DC bands are steep (Figure~\ref{surface_emissivities}, lower panel), their predicted luminosities are largely insensitive to the particular choice of the uncertain outer radius of the BLR, while their characteristic emitting and responding radii (Figure~\ref{dc_halfsolar}) scale roughly as the square-root in the size of the outer radius. Finally, we also point out that the emissivity- and responsivity-weighted radii for the diffuse continuum bands at all UV-optical wavelengths are significantly smaller (factor of $\approx$2) than that predicted for H$\beta$, suggesting that H$\beta$ {\em is not}, as is commonly assumed, a good proxy for the Balmer continuum delays. 

\subsection{Some effects of model assumptions to the predicted BLR diffuse continuum flux and delay spectra}

In Figure~\ref{dc_solar} we show how differences in the model assumptions, and in particular, the choice of gas hydrogen column density and range in hydrogen gas densities affect the diffuse continuum spectrum steady-state luminosities and delays. We explore column densities in the range $\log N_{\rm H} (\rm { cm}^{-2}) = 22.5,23,23.5$ (red, black, and blue solid lines respectively), for our fiducial gas density distribution $\log n_{\rm H} ({\rm cm}^{-3})= 8-12$ (KG00). We also explore a range of gas density distributions spanning $\log n_{\rm H} ({\rm cm}^{-3})= 8-11, 8-12, 8-13$ (dot-dashed, solid, and dashed lines respectively), for our fiducial fixed cloud column density $\log N_{\rm H} (\rm {cm}^{-2})=23$ (KG00). In order to illustrate the direct dependencies on the choice of gas density distribution and cloud column density, we do not re-optimize the BLR geometry (e.g., outer boundary or cloud radial covering fraction distribution) in order to try to refit the observed emission-line strengths.

Figure~\ref{dc_solar} (upper panel, solid lines) illustrates that increasing the cloud column density to higher values {\it increases} the luminosity of the diffuse continuum bands {\it at all wavelengths}, although the spectral shape remains largely unchanged. Most of this extra emission arises from higher column density clouds of higher ionisation parameters. However, to first order the predicted strengths of the major emission lines will respond similarly to changes in the distribution of cloud column densities, which then would result in corresponding changes to the model's overall cloud covering fraction of the source for the same geometry. Therefore, mainly {\em differential} effects between the diffuse continuum and the emission line strengths due to changes in the cloud column density distribution will appear in a resulting predicted model spectrum. Thus, {\em the predicted luminosity dependencies on cloud column density shown in the top panel of Figure~\ref{dc_solar} are exaggerated} relative to models that are constrained by the strengths of the strong emission lines. The two most important such differential effects are due to scattering, both electron scattering and that by neutral hydrogen represented by the strong Rayleigh scattering feature centered on 1216\AA\/. Both become increasingly important in higher column density clouds. 

Clouds with larger neutral hydrogen column densities, typically those with smaller ionisation parameters, will produce broader Rayleigh scattering features. In contrast clouds with larger ionized column densities will generally produce much greyer reflections of the incident continuum due to electron scattering, modulated by bound-free opacity edges (see Korista \& Ferland 1998). An increased electron scattering contribution will thus tend to flatten the UV-optical diffuse continuum flux spectrum emanating from the BLR, and reduce the flux contrast across the bound-free emission jumps. The predicted neutral hydrogen scattering spectral feature places constraints on the amount of neutral hydrogen present (on average) within the BLR\footnote{KG01 noted that their model, while not constrained to do so, predicts this neutral hydrogen reflection feature to lie within the observed Ly$\alpha$ broad-emission line profile in NGC~5548, and may account for its apparently very broad wings.}. However, we caution that physical constraints based upon scattered light are much more emitter/observer geometry dependent than those from the thermal DC emission. Referring now to the middle and lower panels (solid lines), the presence of higher column density clouds at smaller BLR radii {\it reduces\/} $R_{\rm ew}$ and $R_{\rm rw}$ at all wavelengths, while maintaining the overall shape of the delay-spectrum. These characteristic radii and the resulting wavelength-dependent delays (lags) due to variability in the ionising continuum are, of course, insensitive to the overall covering fraction for {\em a fixed geometry}.

In contrast to the luminosity dependence on cloud column density distribution, the dependence in gas density distribution is a largely differential one with respect to the predicted strengths of the major emission lines, because for increasing gas densities the clouds gradually become stronger continuum emitters at the expense of the emitting efficiencies of the major emission lines. In the upper panel of Figure~\ref{dc_solar}, we show that for a fixed cloud total hydrogen column density, increasing the range in hydrogen gas density toward higher densities also acts to increase the DC band luminosities (compare dashed, solid, dot-dashed black lines). Figure~\ref{plot_phi_nh} indicates that the diffuse continuum bands become more emissive at both higher gas densities and hydrogen ionizing photon fluxes, i.e., at smaller BLR radii (see also KG01; Lawther et~al.\ 2018). Thus, cloud gas density distributions that peak toward higher densities tend to increase the luminosity of the DC bands, while at the same time reducing their emissivity- and responsivity-weighted radii. We illustrate the latter in the middle and lower panels of Figure~\ref{dc_solar} (compare dot-dashed, solid, dashed black lines). The inclusion of higher gas densities also diminishes the prominences of the Balmer and Paschen jumps in the wavelength-dependent emissivity- and responsivity-weighted radii, and this would be reflected in the delay spectra as well. 

\subsection{The DC delay signature from BLR clouds}

The steady-state values described above are rarely (if at all) realized in practice, since measured delays depend also on the amplitude and characteristic timescale of the driving ionizing continuum (e.g., Goad \& Korista 2014). Therefore, in order to estimate a representative delay signature for the DC bands, we must first drive our steady-state model with a driving continuum light-curve, one that is broadly representative of {\em the variability amplitude and characteristic timescale\/} of the driving ionizing continuum in this source.

\subsubsection{A proxy for the driving ionising continuum light curve}

\begin{figure}
\includegraphics[width=\columnwidth]{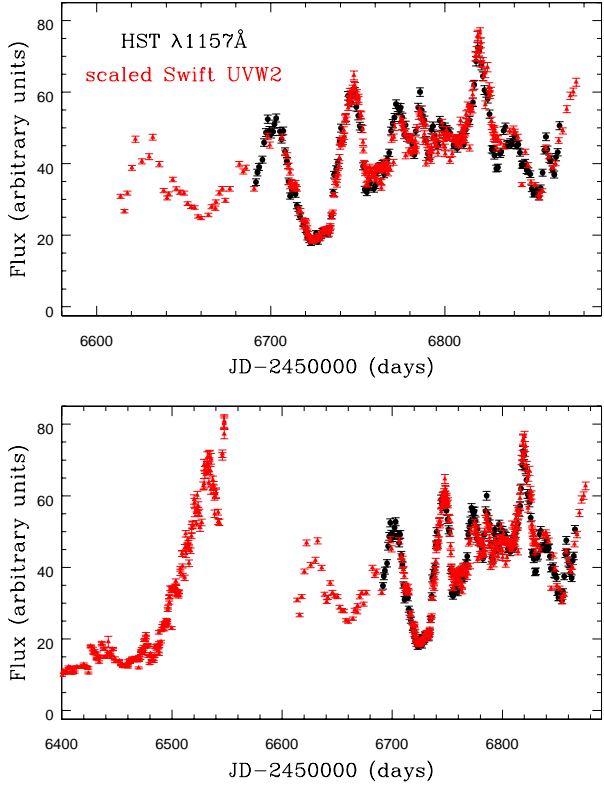}
\caption{
Upper panel -- {\it HST\/} $\lambda$1157\AA\/ continuum band from AGN~STORM (black points). Scaled {\it Swift\/} UVW2 band data taken as part of AGN~STORM (red points). Lower panel -- extended version showing earlier {\it Swift\/} UVW2 data (red points), scaled in the same fashion. Note the significant continuum event $\approx$ 140 days prior to the start of the \textit{HST} campaign.
}
\label{plot_uvw2}
\end{figure}

We begin by adopting the shortest-wavelength available emission-line free continuum light-curve observed during the AGN~STORM campaign, that at $\lambda$1157\AA\/ (observed frame), as a proxy for our driving ionising continuum. This band lies closest in wavelength to the ionising continuum, displays larger amplitude variability than the longer wavelength bands, and as shown by our steady state model (Figure~\ref{dc_halfsolar},\ref{dc_solar}), is among the least contaminated by diffuse continuum emission from the BLR. As our model BLR is large (with a radial extent of 140 light days), and we wish to investigate the response of the {\it whole\/} BLR (spanning delays of 280~days) over the full duration of the {\it HST\/} campaign, we extend the driving light-curve to epochs prior to the start of the {\it HST\/} campaign using {\it Swift\/} UVW2 observations. The {\it Swift\/} UVW2 data have been scaled so that data taken contemporaneously with {\it HST\/}, have the same mean flux and variability amplitude. We ignore here any additional temporal smoothing between the 1157\AA\/ and UVW2 bands which may arise due to differences in the size of the disk emitting regions where these bands originate. We use the scaled {\it Swift\/} UVW2 data only for those epochs prior to the start of the {\it HST\/} campaign, and the {\it HST\/} $\lambda$1157\AA\/ continuum light-curve otherwise. Importantly, the extended light-curve includes a significant continuum event starting $\approx$200 days prior to the start of the {\it HST\/} campaign, and reaching a measured peak flux $\approx$60 days or so later, that exceeds that found during the entire AGN~STORM campaign (Figure~\ref{plot_uvw2}). If the BLR in NGC~5548 is as large as $\sim$100 light-days, then the presence of this continuum event may in part explain the requirement for a large amplitude slowly varying background component at early times in the more commonly employed reverberation mapping recovery techniques (Horne et~al.\ 2019 {\sc paper ix}, in prep.).

\onecolumn
\begin{table}
\begin{center}
\caption{KG00 BLR model predicted CCF lags and correlation coefficients for the 1989 {\em IUE}, 1993 {\em HST}, and 2014 AGN STORM campaigns for NGC~5548}
\label{tab2}
\begin{tabular}{lrrrrrrrrr} \hline 
Line ID  &\multicolumn{3}{c}{IUE 1989} & \multicolumn{3}{c}{HST 1993} & \multicolumn{3}{c}{AGN STORM} \\
         & $\tau_{\rm peak}$ & $\tau_{\rm cent}^{\dagger}$ & Corr & $\tau_{\rm peak}$ & $\tau_{\rm cent}^{\dagger}$ & Corr & $\tau_{\rm peak}$ & $\tau_{\rm cent}^{\dagger}$ & Corr \\
         & \multicolumn{2}{c}{(days)} & coeff. & \multicolumn{2}{c}{(days)} & coeff. & \multicolumn{2}{c}{(days)} & coeff. \\ \hline
Ly$\alpha$  $\lambda$1216  & 10.5 & 11.2 & 0.93 &  8.4 & 10.0 & 0.91 & 6.3 & 7.5 & 0.78\\ 
Si~{\sc iv} $\lambda$1397 & 	10.7	& 11.6	   & 0.94	  & 9.4	 & 10.8		& 0.92	   &  7.5 & 8.7 & 0.80	\\
O~{\sc iv}] $\lambda$1402 & 	10.1	& 10.4	   & 0.95	  & 8.3	 & 9.8		& 0.93	   &  6.6 & 7.7 & 0.83	\\
C~{\sc iv}  $\lambda$1549 & 11.0 & 12.2 & 0.93 &  9.5 & 10.9 & 0.91 &  7.5 & 8.8 & 0.76\\
He~{\sc ii} $\lambda$1640 &  6.0 &  6.5 & 0.97 &  4.6 &  6.0 & 0.95 &  4.2 & 4.7 & 0.90\\
O~{\sc iii}] $\lambda$1663& 	14.8	& 16.8	   & 0.90	  & 14.3	 & 13.7		& 0.91	   & 10.5 & 12.6 & 0.69	  \\
Mg~{\sc ii} $\lambda$2798 & 25.8 & 26.1 & 0.78 & 19.5 & 18.9 & 0.92 & $^{\ddagger}$44.9 & $^{\ddagger}$33.8 & 0.55\\ 
H$\beta$ $\lambda$4861    & 14.6 & 16.4 & 0.86 & 14.3 & 12.7 & 0.90 &  7.4 & 11.4 & 0.64\\
\hline
\end{tabular}
\newline
\end{center}
\noindent $^{\dagger}$ CCF centroids for the broad emission-lines are measured at 80\% maximum.\newline
\noindent $^{\ddagger}$ The predicted Mg~{\sc ii} delays in the AGN STORM campaign are influenced by the presence of the strong continuum event starting $\approx$140 days before the beginning of the {\it HST\/} campaign. Its peak and centroid lags drop by factors of $\approx$2, if this event is excluded.

\end{table}
\twocolumn


\subsubsection{Simulated wavelength-dependent delays and flux variability characteristics of the diffuse continuum}

\begin{figure}
\includegraphics[width=\columnwidth]{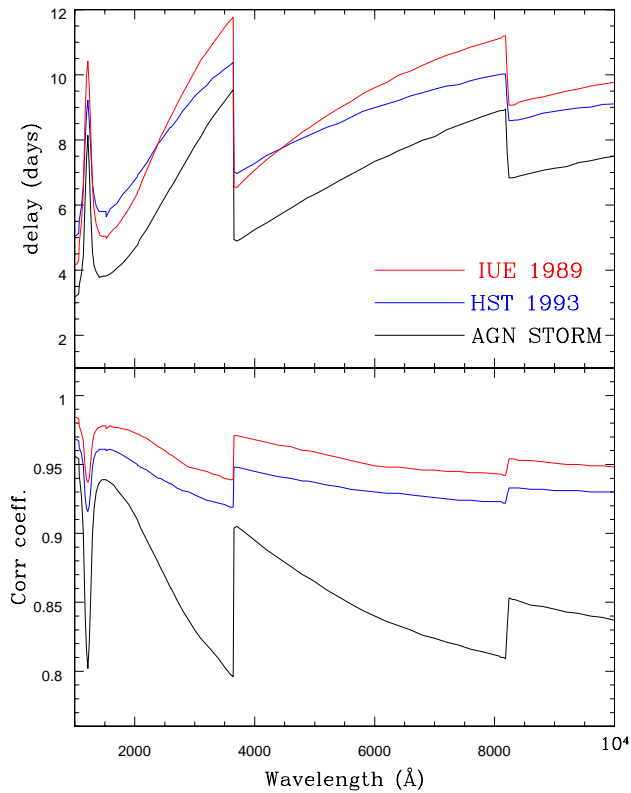}
\caption{Upper panel -- the measured delay (CCF centroid at 0.5~$r_{\rm max}$) as determined from a cross-correlation of a driving continuum light-curve and the wavelength-dependent diffuse continuum contribution arising from the BLR. We use driving continuum light curves from the three extensive monitoring campaigns on NGC~5548; \textit{IUE}~1989 (red), \textit{HST}~1993 (blue), and AGN~STORM (black). Lower panel -- The corresponding peak in the cross-correlation coefficient, in the absence of noise.}
\label{ccf_all}
\end{figure}

We drive our model BLR using our proxy driving continuum light-curve to produce broad emission-line and diffuse continuum light-curves. These we then cross-correlate with the driver to measure their delay (peak and centroid of the cross-correlation function, hereafter CCF) using the CCF implementation of White \& Peterson (1994), interpolating on both light-curves. CCF centroids for the DC bands have been conservatively measured above 50\% $r_{\rm max}$, the maximum correlation coefficient. The measured delays for the DC component only are shown in Figure~\ref{ccf_all}, upper panel. In Figure~\ref{ccf_all}, lower panel, we show the corresponding peak correlation coefficients from cross-correlating the driver with each DC band. Reported delays represent those that would be measured if the DC component could be {\em cleanly isolated from the underlying continuum\/}, and thus represent {\it an upper bound\/} to the DC signature in disk reverberation mapping experiments. Predicted broad emission-line delays (CCF peak, and centroid, the latter measured at 80\% maximum) for our model are presented in Table~\ref{tab2} for comparison with published values of this source for the 2014 AGN~STORM campaign (De~Rosa et~al.\ 2015; Pei et~al.\ 2017). The predicted DC and broad emission-line delays are significantly smaller than that inferred from their steady-state responsivity-weighted radii. This is a consequence of the short characteristic variability timescale, $T_{\rm char}$, of the driving continuum relative to $2R_{\rm rw}/c$ (see e.g., Goad \& Korista 2015), and this is particularly true during the AGN~STORM campaign. We illustrate this further by comparing the DC and broad emission-line delays presented in Figure~\ref{ccf_all} and Table~\ref{tab2} for the AGN~STORM campaign, with those predicted for two alternate driving continuum light-curves, measured at the slightly longer wavelength of $\lambda$1356\AA\/ taken from the 1989 {\it IUE\/} and 1993 {\it HST\/} NGC~5548 monitoring campaigns (Clavel et~al.\ 1991; Peterson et~al.\ 1991; Korista et~al.\ 1995). The continuum light-curves for these three campaigns, differ in both sampling rate, and duration. More importantly their variability behavior differs significantly, despite having similar mean continuum fluxes.

\begin{figure}
\includegraphics[width=6.5cm, angle=270]{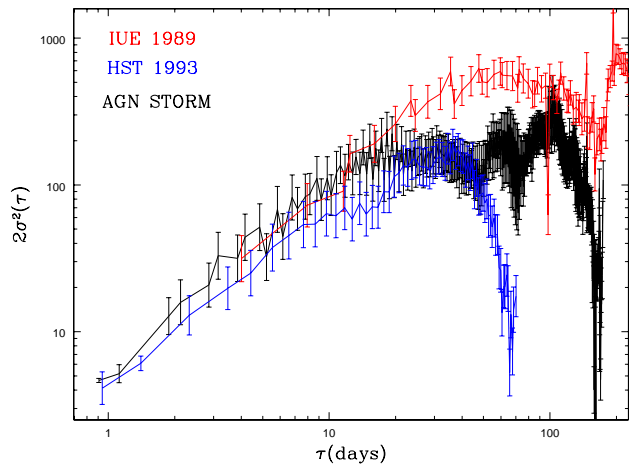}
\caption{
Structure functions for the $\lambda$1350\AA\/ continuum band for the 3 previous space-based reverberation mapping campaigns on NGC~5548. The break timescale for the 1993 and 2014 \textit{HST} campaigns are significantly shorter than that found for the 1989 \textit{IUE} campaign ($\approx$ 40 days), and display significantly less variability power on longer time scales. the power-law index on intermediate time scales is $\approx$1.
}
\label{plot_sf}
\end{figure}

A structure function analysis of the $\lambda$1356\AA\/ UV continuum light-curves for all three campaigns, indicate differences in their characteristic variability timescales and total power (Figure~\ref{plot_sf}). In brief, structure functions are characterised by two flat sections, one on short timescales corresponding to twice the variance of the noise, the other on longer timescales corresponding to twice the variance of the light-curve (e.g., Collier 2001). These are joined on intermediate timescales, by a power-law, the index of which describes the nature of the variability process. The 1989 {\it IUE\/} campaign has the largest power at long timescales ($T_{\rm char} \approx 40$~days, Collier 2001). For the 1993 {\it HST\/} campaign, $T_{\rm char}\approx 20-30$~days. For the AGN~STORM campaign the structure function flattens to twice the variance on timescales of $<20$~days (similar to the FWHM of the continuum auto-correlation function $\approx$12~days for this campaign)\footnote{These values may be compared with a characteristic timescale for the optical continuum variations of $\approx$200~days estimated by Kelly et~al.\ (2009), based on an analysis of $\approx$13~years of AGN~Watch optical monitoring data on NGC~5548.}. The wavelength-dependent DC delays predicted using these alternate drivers are presented in the upper panel of Figure~\ref{ccf_all} (\textit{IUE} 1989, red solid line, and \textit{HST} 1993, blue solid line), along with that predicted for the AGN~STORM campaign (solid black line)\footnote{We note that for the purposes of the present study, we do not account for the anomalous behaviour exhibited by the broad emission-lines reported in Goad et al.\ 2016 (AGN STORM paper~{\sc iv}), and which indicate that the UV-optical continuum may be a poor proxy for the driving EUV continuum during the period of anomalous behaviour.}.

The DC delay spectra for these three campaigns, though broadly similar in shape (the DC from the {\em HST} 1993 campaign appears somewhat flatter), differ in amplitude. The DC from the {\em HST} 1993 campaign shows the smallest spread in delays, the {\em IUE} 1989 campaign, the largest. Also shown (lower panel) are the peak correlation coefficients as a function of wavelength. In general the correlation coefficient is anti-correlated with the delay signature (i.e., the larger the delay the weaker the correlation coefficient). However, a comparison between the measured peak correlation coefficients at each wavelength between campaigns, indicates that on average the peak correlation coefficient is weakest in the AGN STORM campaign and strongest in the {\em IUE} campaign. This, arises because the continuum variations during the AGN STORM campaign, low amplitude and short timescale (e.g., Sun et al.\ 2018), are less effective at driving a response in the DC bands, which preferentially form at radii $R_{\rm ew} > cT_{\rm char}$.
%
%

\begin{figure}
\includegraphics[width=\columnwidth]{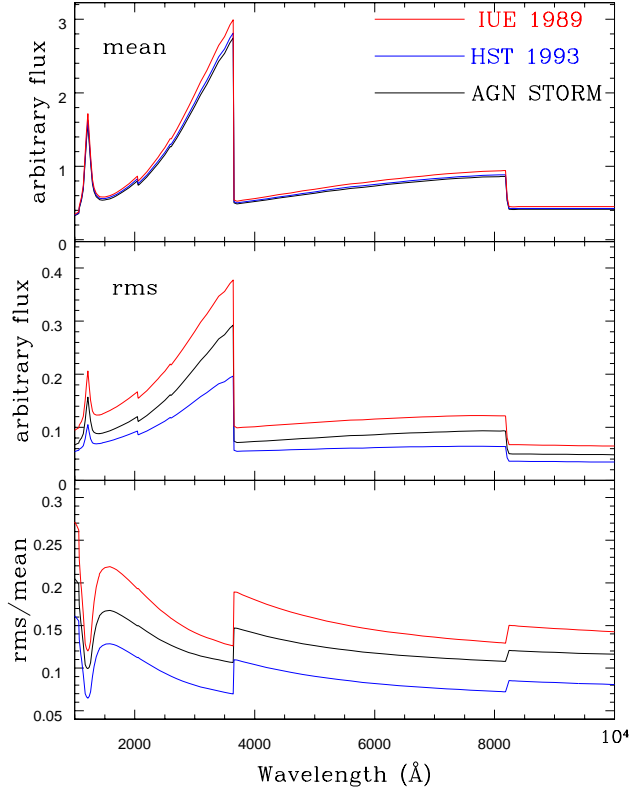}
\caption{Upper panel -- The time-averaged flux spectrum for the KG00 model of NGC~5548.  Colors denote the {\em IUE} 1989 (red), {\em HST} 1993 (blue), and AGN STORM (black) campaigns. Middle panel -- the root mean square (rms) variation as a function of wavelength determined from the diffuse continuum band light-curves. Lower panel -- the fractional variation (rms/mean) as a function of wavelength.}
\label{mean_rms}
\end{figure}

The wavelength-dependent DC light-curve mean fluxes, root mean square (rms) variation and their ratio (rms/mean) for each campaign are shown in Figure~\ref{mean_rms} (upper and lower panels respectively). The nearly identical DC time-averaged flux spectra (top panel) is mainly a reflection of very similar mean ionising continuum luminosities in all 3 campaigns. It is evident that the {\it same BLR} can produce DC light-curves whose response amplitudes (middle and lower panels) and delays and correlation coefficients (Figure~\ref{ccf_all}) differ according to the amplitude and characteristic variability timescale of the driving continuum light-curve. The smallest amplitude DC variations are found for the HST 1993 campaign driver, which has the smallest amplitude variations (rms/mean = 0.193), consistent with its low overall power on long timescales. The largest DC variations are driven by the 1989 {\em IUE} campaign driver (rms/mean = 0.323).
The {\em IUE} 1989 driving continuum produces on average the largest delay signature ($8.5\pm 2.0$ days averaged over $\lambda\lambda$1000--10000\AA), and the AGN~STORM campaign produces the shortest ($6.5\pm 1.7$~days, averaged over the same wavelength range). Similar behavior in the response amplitude and delays resulting from differences in the driving continuum are also exhibited by the broad emission-lines. Note that these average delays for the diffuse continuum is significantly ($\approx$50\%) shorter than that inferred for H$\beta$ using the same continuum driver (Table~\ref{tab2}).  As we mentioned earlier, as regards to a comparison of their emissivity- and responsivity-weighted radii: H$\beta$ and the Balmer continuum do not track one another in their flux variability natures. This is a common misunderstanding in the literature.

\subsection{Dilution of the DC variations by the underlying continuum}

In order to assess the importance of the DC bands to measured continuum variations, we require two additional pieces of information: (i) an estimate of the fractional contribution of the DC light to the total light in each line-free continuum window, and (ii) a model for the underlying incident continuum variations\footnote{The average spectral energy distribution (SED) for the incident UV-optical continuum is the same as previously specified for use in the photoionisation model calculations.}. Only the variable contributions to the continuum light in each band are considered here. For those readers interested in performing detailed spectral decomposition we also present, in Appendix~A, a description of {\it all of the major contributions} to the measured continuum bands, spanning the entire UV-optical--IR continuum.

\begin{figure}
\includegraphics[width=\columnwidth]{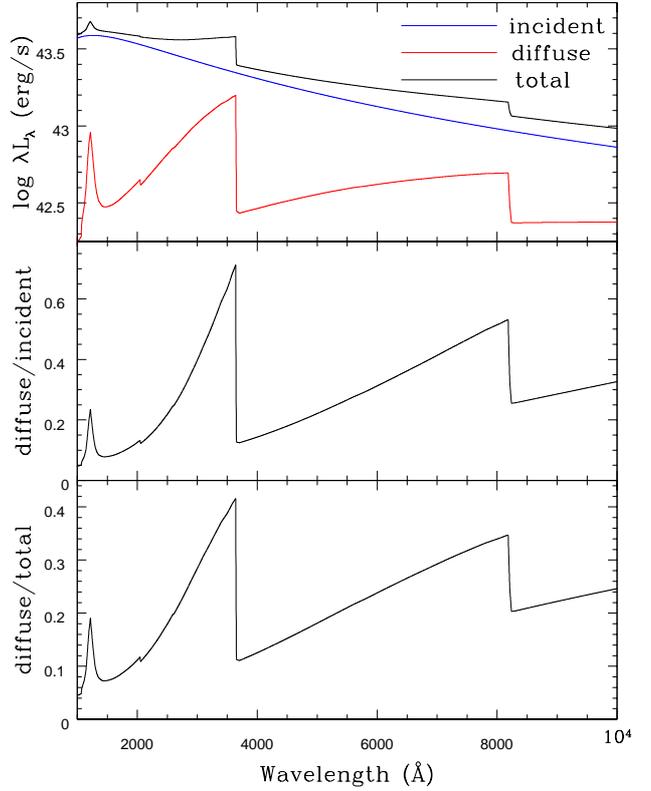}
\caption{Upper panel -- steady-state underlying SED (blue), diffuse continuum emission (red), and their sum (black) as a function of wavelength. Middle panel -- the ratio of the diffuse continuum emission to the incident SED as a function of wavelength. Lower panel -- the ratio of the diffuse continuum emission to the total light (incident + diffuse) as a function of wavelength.}
\label{cont_final}
\end{figure}

In the top panel of Figure~\ref{cont_final} we present the adopted  UV-optical SED (Magdziarz et~al.\ 1998) in the average flux state of the 2014 campaign (blue line) in units of luminosity $\lambda L_{\lambda}$ (erg s$^{-1}$). We note that this UV-optical continuum SED is substantially flatter than the canonical $\lambda L_{\lambda} \propto \lambda^{4/3}$ expected from an accretion disk spectrum peaking in the extreme ultraviolet. This is also true of the UV-optical SED of NGC~5548 derived by Mehdipour et~al.\ (2015). Also shown is the steady-state wavelength-dependent diffuse continuum contribution from our model BLR (red line) {\it in the same units}. The total continuum luminosity, the sum of diffuse$+$incident continuum is indicated by the black line. Note that the sum is significantly flatter than the underlying continuum. As a matter of fact, the inclusion of the diffuse continuum emitted by the BLR is essential when combining with a thermal accretion disk spectrum in order to fit the observed UV-optical-near-IR spectra of AGN.

In the middle panel of Figure~\ref{cont_final} we show the ratio of their luminosities (DC/incident), as a function of wavelength, and in the lower panel the ratio (DC/(DC$+$incident)). The Balmer continuum, and Balmer and Paschen breaks are evident in the combined spectrum (upper panel), clearly illustrating the significance of the DC contribution. The DC contribution rises throughout the Balmer continuum, peaking at Balmer jump, where its strength is $\approx$60\% of the incident continuum light, and $\approx$40\% of the total continuum light in that band. Of course, these values depend on the flux level of the underlying nuclear continuum, presumably from a thermal accretion disk (or one modified by Comptonisation or other reprocessing), which might well be steeper through the UV-optical-near-IR than the SED modeled for NGC~5548, for larger values in $(M\dot{M})^{1/4}$. The contribution of the DC should also track the overall strength of the broad emission line spectrum relative to the underlying continuum; e.g., the equivalent width of broad Ly$\alpha$. We also note that the apparent sharpness of the thresholds in the recombination continua will be reduced relative to that shown here, by smearing due to the BLR velocity field, and significantly by the pile-up of the higher order broad Balmer and Paschen emission lines, plus weak emission from He~{\sc i} and Fe~{\sc{ii}} in the vicinity of the Balmer jump. Additionally, strong UV Fe~{\sc{ii}} emission from the BLR forms a substantial contribution to the spectrum between approximately 2200\AA\/ and 3000\AA\/.

\begin{figure}
\includegraphics[width=6.5cm, angle=270]{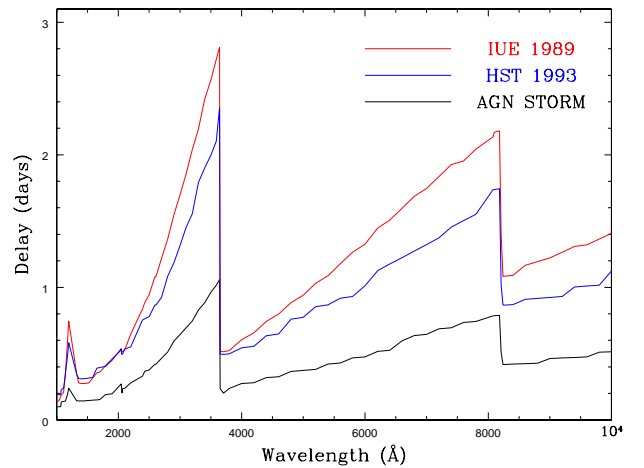}
\caption{The measured wavelength-dependent delays, relative to three different representations of the lag-less driving continuum -- AGN STORM (solid black line), HST 1993 (solid blue line) and {\em IUE} 1989 (solid red line). While the delay signature from the DC contribution is preserved, it is considerably diluted by the presence of the lag-less underlying continuum.}
\label{lag_all}
\end{figure}

We adopt the $\lambda$1157\AA\/ continuum as our proxy driving continuum light curve as before. We further assume that underlying the DC contribution in each wavelength band is a scaled (in flux) version of the driver, which we here {\em assume to be lag-less\/} with respect to the driver. The scale factor between the $\lambda$1157\AA\/ continuum band and longer wavelength bands are taken directly from our model of the underlying SED (Figure~\ref{cont_final}). If the lag-less driving component dominated the luminosity in any continuum band, then we would expect to measure zero delay between the driver and the measured band. Conversely, if the DC were the {\it sole contributor\/} to the measured band, the measured delay should be representative of the delay of the DC component only. Our expectation then, is that {\it for any mixture of lag-less+DC continuum, the measured lag should lie somewhere between these two extremes\/.} Here, we sum the light-curve contributions at each wavelength, scaling the DC component by its fractional contribution to the total continuum luminosity at that wavelength. We then cross-correlate the summed component (representative of a measured continuum band) with the driver as before. The results are presented in Figure~\ref{lag_all}. The measured delays (CCFs centroids at 50\% $r_{\rm max}$) are representative of the delay signature for a lag-less driving continuum contaminated by DC from the BLR. Note that the measured delays are neither lag-less, nor do they equate to that measured for the DC alone (as expected). Instead, the model predicts a DC signature which is broadly representative of that arising from BLR clouds, and with a strong wavelength dependence, but importantly \textit{with significantly smaller delays than predicted for the diffuse continuum alone}. That is, the wavelength-dependent DC delay signature is diluted by the lag-less underlying continuum. Repeating these simulations for the three driving light-curves, described above, again indicates that measured delays within the continuum bands additionally depend on the temporal nature of the driving continuum. We note that these values for the wavelength-dependent delays are not too dissimilar to values reported in the literature and attributed to a reverberating accretion disk. Of course, these results would be modified somewhat in the presence of an underlying continuum with a wavelength-dependent delay spectrum. We return to this issue in the Discussion.

\subsection{The dependence of the measured delays on the DC flux fraction}

\begin{figure}
\includegraphics[width=6.5cm, angle=270]{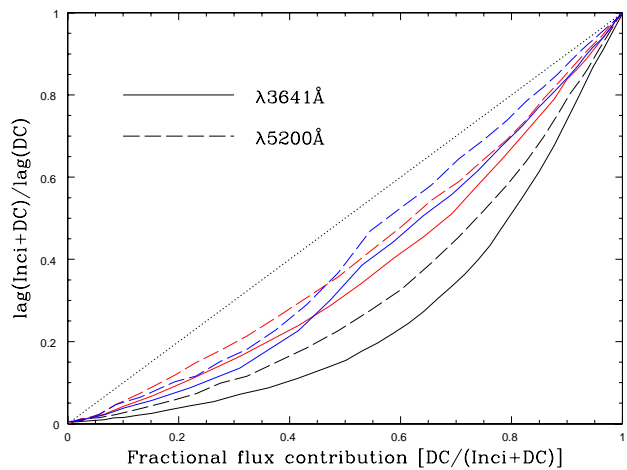}
\caption{Measured delay (CCF centroid at 50\%~$r_{\rm max}$) of the total flux (incident$+$diffuse) continuum light-curve at $\lambda$3641\AA\/ (solid lines), and $\lambda$5200\AA\/ (long-dashed lines), with respect to the driving continuum band at $\lambda$1157\AA\/, plotted as a function of the fractional contribution of the diffuse continuum flux to the total light in that band (DC $+$ incident). Colors indicate the AGN~STORM (black), {\em IUE} 1989 (red), and {\em HST} 1993 (blue) monitoring campaigns. The diagonal grey dotted line indicates a 1:1 relation, as adopted in KG01.}
\label{grow}
\end{figure}

An obvious question to ask, and of immediate interest to disk reverberation mapping campaigns is: ``How does the measured delay depend on the fractional contribution of the diffuse continuum light to the total light in that band?'' Since the delay spectrum resembles the diffuse continuum component, to first order the measured delay must be proportional to it's fractional contribution, as first suggested in KG01. To quantify this further, we choose two representative continuum bands, $\lambda$3641\AA\/ and $\lambda$5200\AA\/ and incrementally adjust their fractional contribution to the measured light in that band. We then proceed as before, driving our model with a lag-less driving continuum, to produce model continuum light-curves, from which we can measure their delays relative to the driver. Figure~\ref{grow} presents the results of this study. Here we plot (on the y-axis) the measured lag as a fraction of the lag measured for the DC continuum alone, versus on the x-axis, the fractional contribution of the DC to the total light (incident + diffuse) in that band. Note that in the limit of zero DC contribution, the lag relative to the driving continuum band is zero (since the underlying continuum is itself lag-less with respect to the driver). Conversely, if the DC dominates the light in the band, the measured delay reproduces that found for the DC alone. However, as illustrated in Figure~\ref{grow}, for intermediate values, the predicted delays, do not generally increase linearly with increasing DC fraction. 

In particular, while the measured delays for the two continuum bands driven by the \textit{IUE}~1989 (red) and \textit{HST}~1993 (blue) light-curves do approximately follow a linear relation, those driven by the AGN~STORM $\lambda$1157\AA\/ continuum (black) follow a relation which approximates to:

\begin{equation}
\tau_{\lambda}(inci+DC) \approx \tau_{\lambda}(DC)\times \frac{(1-A)x}{1-Ax}
\label{recipe}
\end{equation}

\noindent where $x$ is the fractional contribution of the DC to the total light in the band, and $A$ is a constant to be determined. $A=0$ indicates a strictly linear scaling between the flux and lag spectra contributions, as adopted by KG01. A fit to the solid red line in Fig~\ref{grow} yields $A\approx 0.5$, though measured values tend to rise more steeply than this relation as the DC contribution becomes larger than $\approx$50\%. This behaviour may point to a limitation of cross-correlation analyses of multi-variate data. Also, the threshold above which the CCF centroid is measured (50\% $r_{\rm max}$ for the DC bands) was chosen to be conservative, when compared to the value of 80\% $r_{\rm max}$ generally adopted for measuring CCF centroids for the broad emission-lines. Since CCFs are generally skewed toward longer delays, choosing a larger threshold will bias the CCF centroid to be nearer the CCF peak. A fit to the AGN~STORM data (black) yields  A=0.7630 (3641\AA\/, solid), A=0.6485 (5200\AA\/, dashed). We return to this topic in the Discussion.


\subsection{Constraints on the amplitude of the driving continuum flux variations}

Thus far we have assumed that the $\lambda$1157\AA\/ continuum is a suitable proxy for the driving ionising continuum light-curve. However, it has long been known that AGN are generally bluer when brighter (Wamsteker et~al.\ 1990, and references therein); shorter wavelength continuum bands show larger-amplitude and temporally-sharper flux variations. This is likely a natural manifestation of higher energy photons preferentially forming deeper in the gravitational potential well of the supermassive black hole. KG01 suggested that the bluer when brighter effect {\em could in part} be explained by contaminating time-variable diffuse continuum emission from the BLR, and we revisit this further, below.  While the bulk of the ionising continuum cannot, in general, be observed directly, its properties, in particular its variability amplitude, may be inferred from observations of the variability behavior of the observable continuum bands (via extrapolation to shorter wavelengths), as well as that of photon counting broad emission-lines, e.g., He~{\sc ii} $\lambda$1640.

\subsubsection{Estimating the amplitude of the driving continuum}
\begin{figure}
\includegraphics[width=6.5cm, angle=270]{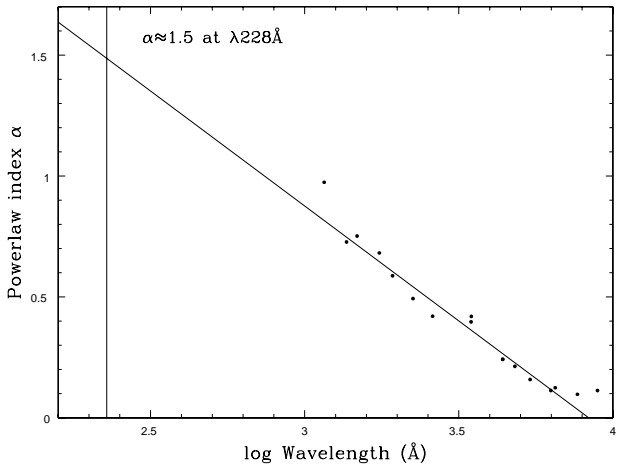}
\caption{An estimate for the variability amplitude of the driving continuum at shorter wavelengths. Values of $\alpha$ have been determined from a fit to the relation  $\log_{\rm 10} F(\lambda) = \alpha \log_{\rm 10}F(\lambda 1157) + C$\,.   }
\label{plot_scale}
\end{figure}

Analysis of the root mean square (rms) variability amplitude of 19 continuum bands presented in Fausnaugh et~al.\ (2016), for NGC~5548, and {\em after removal of the host galaxy contribution}, indicate that the continuum variability amplitude decreases with increasing wavelength. We show this in Figure~\ref{plot_scale}. This signal is broadly consistent with that expected from X-ray reprocessing of the continuum in a spatially extended disk, and as we have shown above, is also consistent with that expected to arise from reprocessing of the EUV continuum in the spatially extended BLR. By extrapolating the wavelength dependence of the rms relation to shorter wavelengths we can obtain a crude estimate for the variability amplitude of the EUV continuum. This indicates that the rms amplitude at EUV wavelengths (228\AA\/) is $\approx 0.37$.

Alternatively, we can derive a more accurate estimate for the amplitude of variability at shorter wavelengths by estimating the power-law index $\alpha$ in the relation $\log_{\rm 10} F(\lambda) = \alpha \log_{\rm 10}F(\lambda 1157) + C$. We have applied this scaling relation to each of the longer wavelength UV-optical continuum bands. In Figure~\ref{plot_scale}, we plot the measured values of $\alpha$ versus logarithmic wavelength. Extrapolating a fit to this relation to shorter wavelengths yields an estimate for $\alpha$ necessary to map the observed continuum variations at $\lambda$1157\AA\/ into the EUV regime. Using this procedure, we derive a best-guess estimate for $\alpha$ at 228\AA\/ of $\approx$1.5. The value of rms/mean for the AGN~STORM $\lambda$1157\AA\/ continuum after scaling by this factor is 0.366, similar to the value obtained above from extrapolation of the rms relation to shorter wavelengths. While uncertainty remains as to whether such an extrapolation is appropriate (and can not extend to arbitrarily large photon energies), we note that these findings are consistent with those of Marshall et~al.\ (1997). They report on the variability behavior of NGC~5548 in the extreme ultraviolet ($\lambda\lambda$70--100\AA\/) over a 2 month period with the {\em Extreme Ultraviolet Explorer} (hereafter, \textit{EUVE}), monitored as part of 1993 multi-wavelength space and ground-based monitoring campaign on this source (Korista et~al.\ 1995). \textit{EUVE} observations simultaneous with observations at UV (\textit{HST} and \textit{IUE}) wavelengths, indicate a factor $\approx$1.7 larger variability amplitude than those at UV wavelengths ($\lambda$1350\AA\/).
%

\begin{figure}
\includegraphics[width=\columnwidth]{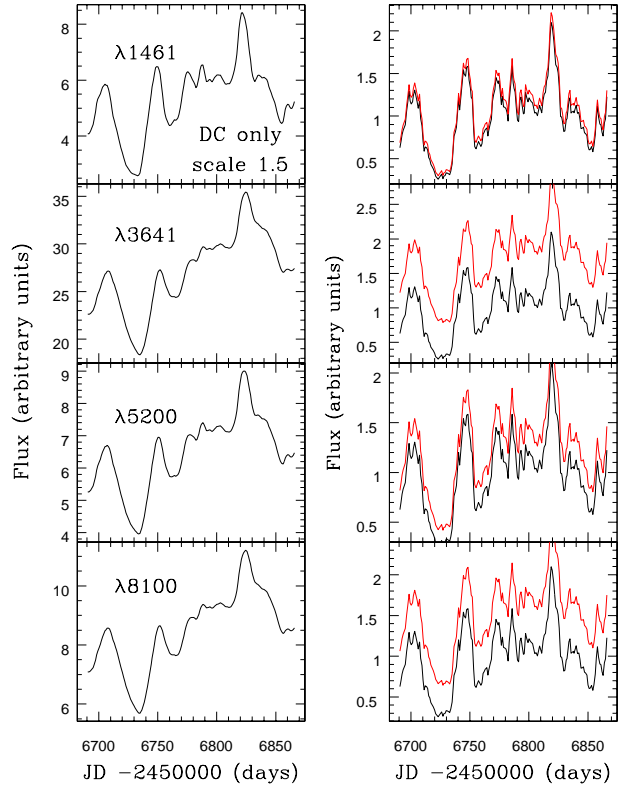}
\caption{Left -- representative diffuse continuum light-curves constructed using our amplitude-scaled proxy driving continuum ($\alpha=1.5$, see text for details). Right -- the incident continuum in normalized units (black) and the sum of incident $+$ diffuse continuum (red), for the same continuum bands and in the same units. }
\label{dc_lcurves}
\end{figure}

\begin{figure}
\includegraphics[width=6.5cm, angle=270]{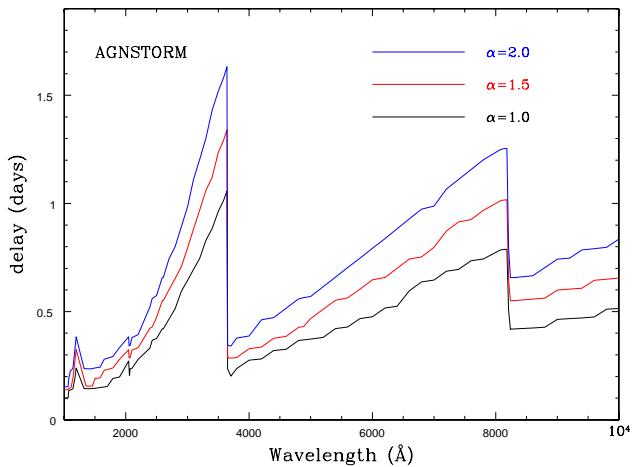}
\caption{
Sum of incident $+$ diffuse continuum correlated with respect to the $\lambda$1157\AA\/ continuum band. The $\lambda$1157\AA\/ driver is scaled according to $\log_{10} F({\rm driver}) = \alpha \log_{\rm 10} F(\lambda 1157) + C\,$, with $\alpha=1.0$ (black) $\alpha=1.5$ (red), and $\alpha=2.0$ (blue).
}
\label{lag_scale}
\end{figure}

\subsubsection{The effect of enhanced amplitude continuum variations on the measured delays}

Increased amplitude variations in the driving continuum will induce larger amplitude variations in the DC bands. This alone {\it will not affect\/} the measured DC delays. However, enhanced amplitude DC variations will {\it alter the fractional contribution of the DC to the total light in that band\/} (relative to that predicted by a lower amplitude driver) {\em in a time-dependent manner}. And as can be seen from Figure~\ref{grow}, a change in the DC fractional contribution to the total light in the band does affect the measured inter-band continuum delays, which are here referenced to the measured continuum at 1157\AA\/.

We demonstrate this effect by repeating our simulations described in \S2.6 for two additional drivers with enhanced amplitude variations ($\alpha=1.5$, and $\alpha=2.0$). As before, we add the (un-scaled) underlying continuum light-curve and diffuse continuum light-curves together according to their fractional contributions to the total light in that band. We show examples of this procedure for 4 continuum bands, adopting $\alpha=1.5$ in  Figure~\ref{dc_lcurves}. We then cross-correlate the summed light-curves for each band with the observed 1157\AA\/ continuum light-curve, and measure the CCF peak, and centroid (for correlation coefficients $>$ 50\%  $r_{\rm max}$). In Figure~\ref{lag_scale} we compare the wavelength-dependent delays for 3 different driving continuum amplitudes. We confirm that an enhanced driving continuum amplitude acts to increase the DC contribution, and thereby enhances the measured delays for all wavebands.

\subsubsection{Bluer when brighter}
\begin{figure}
\includegraphics[width=\columnwidth]{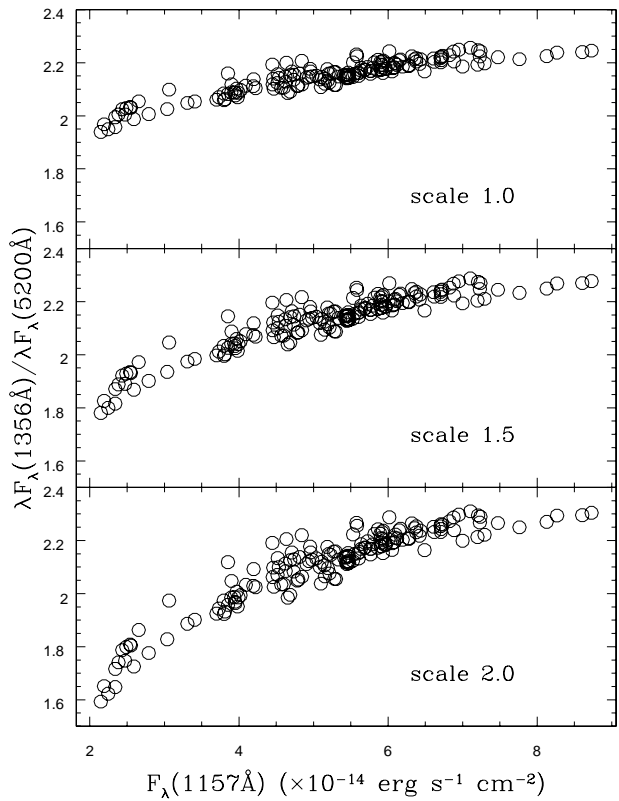}
\caption{
The ratio of the integrated flux ($\lambda F_{\lambda}$) in two representative continuum bands $\lambda$1356\AA\/ and $\lambda$5200\AA\/ versus the observed AGN~STORM HST $\lambda$1157\AA\/ continuum light-curve. The flux ratio increases as the continuum flux increases, and contributes in a minor yet significant way to the well-known ``bluer when brighter'' effect observed in AGN. Scale factors, $\alpha$, are as in Figure~14 (see text for details).} 
\label{bluer}
\end{figure}

As noted by KG01, a time and wavelength variable DC contribution can explain a significant fraction of the ``bluer when brighter'' effect observed in AGN continuum variability studies. This we illustrate in Figure~\ref{bluer} where we plot the ratio of the 1356\AA\/ to 5200\AA\/ fluxes (each containing the underlying continuum plus the DC) versus the specific flux at 1157\AA\/ (assumed to be proportional to the driving ionising continuum flux). Here, our model underlying continuum {\em does not} change its spectral shape, so if there were no color variation, this UV -- optical flux ratio would be independent of the 1157\AA\/ continuum flux. Instead, this ratio increases with increasing flux, and at a rate which is faster for lower central source continuum fluxes. The magnitude of this effect is also notably larger for a larger amplitude driving continuum. For example, compare the upper and lower panels of Figure~\ref{bluer}, increasing from a 10\% to a 50\% effect over a factor of $\sim$5 in the observable proxy for the driving ionising continuum flux.

\section{Discussion}

State-of-the-art reverberation mapping experiments are now revealing significant wavelength-dependent continuum time-delays within the X-ray-UV-optical-IR continuum bands in several nearby AGN (e.g., Edelson et al.\ 2015; Fausnaugh et al.\ 2016; Cackett et~al.\ 2018, Edelson et~al.\ 2019). Whether the measured continuum inter-band delay signature is associated with reprocessing of X-ray photons in the disk remains unclear (Chelouche et~al.\ 2019). The factor few larger than expected delays remains problematic and may point to alternate disk geometries (Dexter \& Agol 2011; Hall et~al.\ 2018). Furthermore, large uncertainties in the relative contribution to the delay signature of additional reprocessing sites, for example, the proposed EUV torus (Gardner \& Done 2017) and the known BLR (Korista \& Goad 2001; Lawther et~al.\ 2018) are yet to be resolved. Here we have attempted to estimate the likely contribution to the delay signature of DC emission {\em from the same gas emitting the broad emission-lines}, using the well-studied nearby Seyfert~1.5 galaxy, NGC~5548, as a point of reference. Our approach is holistic, in that our model must first match the observed broad emission-line strengths, which provide the all-important additional constraints upon the gas physics, and {\em only once this condition is satisfied\/} do we compute the DC flux and delay spectra.

In what follows we discuss two separate considerations of the role of geometry in determining the flux and delay spectra of the BLR's diffuse continuum, as well as a means of estimating the diffuse continuum's contribution to the measured inter-band continuum delays. 

\subsection{The effects of large-scale geometry on the DC flux and delay spectra}

The gas responsible for the broad emission lines contributes significantly to the UV-optical continuum flux. Forming at relatively small radii and varying with an approximately linear response amplitude together suggest that the DC bands are {\em less sensitive to uncertainties in the BLR geometry\/} than many of the broad emission lines, and should be accurate tracers of the driving continuum variations at ionising energies. 
Similar arguments can also be applied to the broad UV-optical He~{\sc ii} emission lines. Refer to Figure~\ref{surface_emissivities}. 

To test the sensitivity of the DC delays to the assumed geometry we replace the spherical BLR model of KG00 with the bowl-shaped geometry for NGC~5548 presented in Goad et~al.\ (2012). In brief, in this model the clouds responsible for the broad emission lines occupy a region that approximates the surface of a bowl, for which we see emission only from gas on the observer side of the central plane of symmetry defined by the accretion disk/obscuring torus. This model's BLR spans 1--100 light-days in radius, such that the measured delay at the outer radius, when viewed face-on, is 50 days (corresponding to a scale height (z/r) at the outer radius of 0.5), the top rim of the bowl geometry then approximating the measured delay for the hot dust in this source (Suganuma et~al.\ 2006). The system is then inclined at an angle of 30~degrees with respect to the line of sight, similar to the inclination inferred from dynamical model fits to optical continuum--emission line variability data of NGC~5548 (Pancoast et~al.\ 2014b). In addition to the break in spherical symmetry, which introduces dependencies in inclination, the bowl geometry is smaller in size by a factor of 1.4, and presents emission to the observer from only 1/2 of the 4$\pi$ steradians surrounding the central engine, but has little emitting gas near the observer line of sight for the adopted inclination. Using the same radial surface flux emissivities in the DC continuum bands, we then drive this model with the extended continuum light curve described in \S2.4 (2014 campaign). We find that the steady-state DC flux and delay spectra for our bowl-shaped model are qualitatively similar in shape and magnitude (the latter to within $\approx$ 30\%) to that inferred for the larger, spherical BLR model of KG00. Thus, we confirm that the choice of BLR geometry appears to be of lesser importance in influencing the form of the DC flux and lag spectra than does the BLR gas physics.

\subsection{The Lyman continuum and the question of open vs.\ closed geometries for the BLR}

The standard photoionisation models of broad emission line clouds predict {\em emission} by a substantial Lyman continuum. However, a Lyman continuum has {\em never} been observed in AGN spectra, and as was discussed in KG01 this has been a long-standing problem of photoionisation model predictions (e.g., Carswell \& Ferland 1988). The Lyman continuum must either be significantly suppressed or otherwise largely hidden from view --- both are geometry-dependent. Because the Lyman continuum is ionising, its absence in AGN spectra is perhaps indicative of a particular geometry of the broad emission line gas which does not include clouds in an {\em open} geometry; i.e., one in which all photons emitted by a particular cloud escape the geometry. This has been a standard assumption in published photoionisation models of the BLR. In clouds containing a H-ionisation front, the Lyman continuum is emitted from their illuminated faces, and in open geometries these photons then escape the BLR without further interaction. 

One means of suppressing the Lyman continuum is to invoke a largely {\em closed} emission line geometry. Lyman continuum photons emerging from the illuminated faces of BLR clouds are then typically reprocessed elsewhere in the BLR, rather than escaping from the geometry. The fact that all BLR photoionisation models require the line-emitting geometry to subtend a large solid angle centered on the central continuum source (covering fractions $\approx$30-50\%) to account for the observed emission line strengths, may itself be indicative of a geometry that is not fully open. We have checked for individual photoionisation computations, that adopting a closed geometry results in a dramatic reduction in the escaping {\em ionising} diffuse emission, with only modest changes to the predicted diffuse line and continuum emission at wavelengths longward of the Lyman limit. We note, however, that high-excitation and generally high-ionisation emission lines lying within the Lyman continuum (e.g., N~{\sc iv}, O~{\sc iii}, O~{\sc iv}, O~{\sc v}, and Ne~{\sc viii}) have been identified in AGN spectra (see Shull, Stevans, \& Danforth 2012 and references therein). Thus, the suppression of {\em these} escaping photons cannot be too great. Notably, the high-ionisation and high-excitation natures of these transitions places their origins primarily within high-density gas ($\log {\rm n}_{\rm H} ({\rm cm}^{-3}) \gtrsim 12$) of high incident ionising photon fluxes ($\log \Phi_{\rm H}~({\rm s}^{-1}~{\rm cm}^{-2}) \gtrsim 22$) (see Korista et~al.\ 1997; Moloney \& Shull 2014), corresponding to $r \lesssim 1.3$ light days in NGC~5548. These conditions are atypical of gas responsible for most of the broad emission lines and thus a fair fraction of the Lyman continuum emission from an amalgam of clouds in open geometries. In any case the mystery of the missing Lyman continuum emission remains an unsolved puzzle, but surely one that is telling us something important about the BLR in AGN. 

\subsection{Estimating the DC contribution to measured continuum inter-band delays}

In accretion disk reverberation mapping experiments continuum inter band delays are measured relative to a single reference band, typically one with the shortest wavelength and hence largest amplitude variations, or alternatively, highest S/N. However, the measured light in any approximately emission line-free continuum band at a minimum comprises the variable underlying continuum (which likely shows a wavelength dependence in both amplitude and delay on short timescales), and as we have discussed here, a significant and more slowly varying diffuse continuum emission arising from the BLR. Since neither component is measured in isolation, perhaps only in the presence of a large amplitude short-timescale change in the incident continuum emission can these two components easily be separated. Here, we have estimated the continuum inter-band delays arising from the BLR and which represent a nuisance contaminating component in disk reverberation mapping experiments (e.g., Fausnaugh et~al.\ 2015; Edelson et~al.\ 2015, 2017, 2019; Cackett et~al.\ 2007, 2018). 

We have shown that the $\lambda$-dependent DC flux and delay distributions depend on the distribution in gas density and to a lesser degree the cloud column density (\S2.2, Figure~\ref{surface_emissivities}). The DC delay distribution also depends upon the amplitude and characteristic timescale of the driving continuum (varying by up to 30\% for the model shown here, \S2.4, Figure~\ref{ccf_all}), but as already discussed is found to be relatively weakly dependent on the BLR geometry. The measured continuum inter-band delays depend not only on the DC delay distribution, but also on the fractional contribution of the DC light to the total light in the measured band (\S2.6). Unfortunately, as shown in Figure~\ref{grow} the functional form of this relation is not strictly linear (cf., KG01). Since each continuum band contains contributions from at least two components, a key question then is how does one estimate from observations the DC contribution to the measured delay signature, and thereby reveal the underlying accretion disk delay signature?

In the following, we provide a simple recipe for estimating the DC contribution based on a scaling of our results for NGC~5548. To match the delays appropriate for a particular AGN's luminosity, the diffuse continuum delay spectra from Figure~\ref{ccf_all} should first be scaled by the relative source luminosities ($\rm{L_{AGN}/L_{NGC~5548}}$) using the familiar radius--luminosity relation; $R_{BLR} \propto L^{1/2}$. Since the BLR is large compared to the (predicted) size of the accretion disk responsible for the UV-optical continuum, these scaled delays will represent {\em an upper bound} to the DC delay contribution (under the incorrect assumption that the DC light is the sole contributor to the light in that band). Next, if the DC contribution to the total flux in a given continuum band can be estimated from a fit to the flux spectrum, the DC delay contribution may then be estimated from Figure~\ref{grow}. We note here that it is often standard practice in spectral decompositions to use a scaled spectral template for the Balmer continuum emission (e.g., Fausnaugh et~al.\ 2015), although when used in isolation in combination with a powerlaw continuum, this would indicate a lower limit to the DC contribution. 

If on the other hand, the DC contribution cannot be directly estimated from spectral decomposition, then we suggest using Equation~\ref{recipe}, to fit the measured delays as a function of wavelength, using the scaled template, and optimising over $x$ ($0 \leq x \leq 1$); the DC fractional contribution to the total light in that band), and scale factor $A$, where $x$ and $A$ are free parameters. The constraints are then the measured delays and the spectrum of the source. The fit could also include a model for the disk reverberation signature. 
As long as the wavelength-dependent lags in the underlying continuum are much shorter than those of the DC originating in the BLR, the above recipe should remain useful. A more thorough treatment of this problem will require detailed forward modelling of the variable incident continuum and responding disk and DC components.

We note that in the presence of multiple time-variable signals (e.g., the accretion disk continuum and diffuse continuum from the BLR, and potentially others), each of differing fractional flux contribution, the measured delay within a particular wavelength band represents a {\em lower limit} to the delay of the most slowly varying component, relative to the light associated with driving the observed flux variations.

Finally, in Appendix~A, we explore the potential additional, though generally relatively minor, flux contributions of the continuum emanating from the narrow-emission line region and toroidal obscuring region.

\section{Summary of Findings}

Below we summarize the key results of this work:
\begin{itemize}
\item BLR models which match the observed emission-line luminosities of the strongest UV-optical broad emission-lines also produce significant diffuse continuum (DC) emission. In combination with model thermal accretion disk spectra, this contribution flattens the UV-optical spectrum to better match what is observed in AGN, with significant flux contributions out to $\approx$2 microns, or so.

\item The DC forms over a broad range in radii, and hence delays, and displays a strong wavelength dependence. Delays generally increase with increasing wavelength, but exhibit sharp departures from this general trend in the vicinity of the Balmer and Paschen jumps.

\item The radial surface emissivity distribution $F(r)$, for the DC bands approximates a simple powerlaw of slope $\approx -2$ over much of the BLR (i.e. $F(r) \propto r^{-2}$). This implies a radial responsivity distribution $\eta(r) \approx 1$. Thus, the physics of the diffuse continuum is relatively straightforward, and suggests that the variability behaviour of the DC bands are substantially less dependent on uncertainties in the BLR geometry than that from the optically thick broad emission lines.

\item Inclusion of clouds with higher values in hydrogen gas number and column densities, particularly if concentrated at smaller BLR radii, act to increase the DC contribution to the total light in a given band, while at the same time reducing their typical mean formation radius.

\item The emissivity- and responsivity-weighted radii in the vicinity of the Balmer continuum are significantly (factor of $\approx$2) smaller than that for the more prominent broad Balmer emission-lines. Hence, the H$\beta$ lag is in fact a poor proxy for the Balmer continuum's variability behaviour or its contribution to the continuum band delay.

\item The DC contribution to the measured inter-band continuum delays depends critically upon its flux strength relative to the underlying continuum. However this relation is not a simple (i.e., linear) one. Based on simulations we derive an approximate recipe for correcting the measured delays for contributions from the DC of the BLR.

\item As for the broad emission-lines, the delays for the DC bands, and so their contributions to the measured inter-band continuum delays, also depend on the variability {\em amplitude} and {\em characteristic variability timescale} of the driving continuum. The amplitude of the driving continuum flux variations affects not only our ability to measure a delay (since small amplitude variations will be washed out by an extended reprocessing region), but additionally changes the fractional contribution of the DC light in the measured continuum band, altering its relative importance.

\end{itemize}

\section*{Acknowledgements}
We thank the anonymous referee for providing comments and suggestions which led to improvements in clarity of the work presented here. We thank Gary Ferland for his continued development and support of the photoionisation code, Cloudy. KTK is grateful for the hospitality of the University of Leicester. 





\begin{thebibliography}{99}

\bibitem[Arav et~al.\ (2015)]{ARAV15}
Arav, N., Chamberlain, C., Kriss, G.A., Kaastra, G.A., Cappi, M., et~al.\ 2015, A\&A 577, 37.

\bibitem[Baldwin et~al.\ (1995)]{BAL95}
Baldwin, J., Ferland, G., Korista, K., \& Verner, D.\ 1995, \apj\ 445, L119.

\bibitem[Baskin et~al.\ (2014)]{BLS14}
Baskin, A., Laor, A., \& Stern, J.\ 2014, \mnras\ 438, 604.

\bibitem[Baskin \& Laor (2018)]{BLS18}
Baskin, A., Laor, A.\ 2018, \mnras\ 474, 1970.

\bibitem[Bentz et~al.\ (2010)]{BEN10}
Bentz, M.~C., Horne, K.D., Barth, A.~J., et~al.\ 2010, \apjl\ 720, 46.

\bibitem[Bottorff et~al.\ (1997)]{BOT97}
Bottorff, M., Korista, K.T., Schlosman, I., \& Blandford, R.\ 1997, \apj\ 479, 200.

\bibitem[Bottorff et~al.\ (2002)]{BOT02}
Bottorff, M., Baldwin, J.A., Ferland, G.J., Ferguson, J.W., \& Korista, K.T.\ 2002, \apj\ 581, 932.

\bibitem[Cackett et~al.\ (2007)]{CAC07}
Cackett, E.M., Horne, K.D., \& Winkler, H.\ 2007, \mnras\ 380, 669.  

\bibitem[Cackett et~al.\ (2018)]{CAC18}
Cackett, E.M., Chiang, C-Y., McHardy, I. et~al.\ 2018, \apj\ 857, 53.  

\bibitem[Carswell \& Ferland (1988)]{CF88}
Carswell, R., \& Ferland, G.J.\ 1998, \mnras\ 235, 1121.

\bibitem[Chelouche (2019)]{CD19}
Chelouche, D., Pozo~Nunez, F., \& Kaspi, S.\ 2019, Nature Astronomy, 3, 251.

\bibitem[Chiang \& Murray (1996)]{Chiang96}
Chiang, J., \& Murray, N.\ 1996, \apj\ 466, 704.

\bibitem[Clavel et~al.\ (1991)]{C91}
Clavel, J., Reichert, G.A., Alloin, D., et~al.\ 1991, \apj\ 366, 64. 

\bibitem[Collier (2001)]{COL01}
Collier, S. 2001, \mnras\ 325, 1527.

\bibitem[Dexter \& Agol (2011)]{DA11}
Dexter, J., \& Agol, E.\ 2011, \apj\ 727, 24.

\bibitem[De Rosa et~al.\ (2015)]{STORM01} 
De~Rosa, G., Peterson, B.M., Ely, J., et~al.\ 2015, \apj\ 806, 128, (Paper~{\sc i}).

\bibitem[Edelson et~al.\ (2015)]{STORM02} 
Edelson, R., Gelbord, J.M., Horne, K.D., et~al.\ 2015, \apj\ 806, 129, (Paper~{\sc ii}).

\bibitem[Edelson et~al.\ (2017)]{REd17}
Edelson, R., Gelbord, J., Cackett, E., Connolly, S., Done, C., et al.\ 2017, \apj\ 840, 41.

\bibitem[Edelson et~al.\ (2018)]{REd18}
Edelson, R., Gelbord, J., Cackett, E., Peterson, B.M., Horne, K. et al.\ 2019, \apj\ 870, 123.

\bibitem[Fausnaugh et~al.\ (2016)]{STORM03} 
Fausnaugh, M.M., Denney, K.D., Barth, A.J., Bentz, M.C., Bottorff, M.C. et al.\ 2016, \apj\ 821, 56, ( Paper~{\sc iii}).

\bibitem[Ferland et~al.\ (2017)]{FER17}
Ferland, G.J., Chatzikos, M., Guzm\'an, F., Lykins, M.L., van Hoof, P.A.M. et~al.\ 2017, RMxAA 53, 385.

\bibitem[Fitzpatrick (1999)]{F99}
Fitzpatrick, E.L.\ 1999, PASP 111, 63.

\bibitem[Gardner \& Done (2017)]{GD17}
Gardner, E., \& Done, C.\ 2017, \mnras\ 470, 3591.

\bibitem[Goad et~al.\ (1993)]{GOG93}
Goad, M.R., O'Brien, P.T., \& Gondhalekar, P.M.\ 1993, \mnras\ 263, 149.

\bibitem[Goad et~al.\ (2012)]{GKR12}
Goad, M.R., Korista, K.T., \& Ruff, A.J.\ 2012, \mnras\ 426, 3088.

\bibitem[Goad \& Korista (2014)]{GK14}
Goad, M.R., \& Korista, K.T.\ 2014, \mnras\ 444, 43.

\bibitem[Goad \& Korista (2015)]{GK15}
Goad, M.R., \& Korista, K.T.\ 2015, \mnras\ 453, 3662.

\bibitem[Goad et~al.\ (2016)]{STORM04} 
Goad, M.R., Korista, K.T., De~Rosa, G., Kriss, G.A., Edelson, R., et~al.\ 2016, \apj\ 824, 11, (Paper~{\sc iv}).

\bibitem[Grier et~al.\ (2012)]{GR12}
Grier, C.J., Peterson, B.M., Pogge, R.W., et~al.\ 2012, \apj\ 755, 60.
  
\bibitem[Grier et~al.\ (2013)]{GR13}
Grier, C.J., Peterson, B.M., Horne, K., Bentz, M.C., Pogge, R.W. et~al.\ 2013, \apj\ 764, 47.

\bibitem[Grier et~al.\ (2017)]{GR17}
Grier, C.J., Pancoast, A., Barth, A.J., Fausnaugh, M.M., Brewer, B.J. et~al.\ 2017, \apj\ 849, 146.

\bibitem[Hall et~al.\ (2018)]{HA18}
Hall, P.B., Sarrouh, G.T., \& Horne, K.D.\ 2018, \apj\ 854, 93.
 
\bibitem[Horne et~al.\ (1991)]{KDH91}
Horne, K.D., Welsh, W.F., \& Peterson, B.M.\ 1991, \apj\ 367, 5.

\bibitem[Horne et~al.\ (2019)]{KDH19}
Horne, K., De Rosa, G., Peterson, B.M., Barth, A.J., Ely, J., et~al.\ 2019, \apj\ in prep., {\sc paper ix}

\bibitem[Kaspi \& Netzer (1999)]{KN99}
Kaspi, S., \& Netzer, H.\ 1999, \apj\ 524, 71.

\bibitem[Kaastra et~al.\ (2014)]{KAS14}
Kaastra, J.S., Kriss, G.A., Cappi, M., Mehdipour, M., Petrucci, P.-O., et~al.\ 2014, Sci 345, 64. 

\bibitem[Kelly et~al.\ (2009)]{KEL09}
Kelly, B.C., Bechtold, J, \& Siemiginowska, A.\ 2009, \apj\ 698, 895.

\bibitem[Korista et~al.\ (1995)]{KTK95}
Korista, K.T., Alloin, D., Barr, P., Clavel, J., Cohen, R.D., et~al.\ 1995, \apjs\ 97, 285.

\bibitem[Korista \& Ferland (1998)]{KF98}
Korista, K.T., \& Ferland, G.J.\ 1998, \apj\ 495, 672.

\bibitem[Korista \& Goad (2000)]{KG00}
Korista, K.T., \& Goad, M.R.\ 2000, \apj\ 536, 284 (KG00).

\bibitem[Korista \& Goad (2001)]{KG01}
Korista, K.T., \& Goad, M.R.\ 2001, \apj\ 553, 695 (KG01).

\bibitem[Korista \& Goad (2004)]{KG04}
Korista, K.T., \& Goad, M.R.\ 2004, \apj\ 606, 749.

\bibitem[Koshida et~al.\ (2014)]{KOSH15}
Koshida, S., Minezaki, T., Yoshii, Y., Kobayashi, Y., Sakata, Y., et~al.\ 2014, \apj\ 788, 159.

\bibitem[Kraemer et~al.\ (1998)]{KR98}
Kraemer, S.B. Crenshaw, D. Michael; Filippenko, Alexei V.; Peterson, Bradley M.\ 1998, \apj\ 499, 719.

\bibitem[Kriss et~al.\ (2019)]{KGA19}
Kriss, G.A., De Rosa, G., Peterson, B.M., et~al.\ 2019, \apj\ in press, AGN STORM {\sc paper viii}

\bibitem[Lawther et~al.\ (2018)]{LAW18}
Lawther, D., Goad, M.R., Korista, K.T., Ulrich, O., \& Vestergaard, M.\ 2018, \mnras\ 481, 533.

\bibitem[Landt et~al.\ 2019]{Landt19}
Landt, H., et~al.\ 2019, \mnras\ in press.

\bibitem[Magdziarz et~al.\ (1998)]{MAG98}
Magdziarz, P., Blaes, O.M., Zdziarski, A.A., Johnson, W.N., and Smith, D.A.\ 1998, \mnras\ 301, 179.

\bibitem[Mannucci et~al.\ 1992]{MAN92}
Mannucci, F., Salvati, M., \& Stanga, R.M.\ 1992, \apj\ 394, 98.

\bibitem[Marshall et~al.\ (1997)]{MAR97}
Marshall, H.L., Carone, T.E., Peterson, B.M., Clavel, J., Crenshaw, D.M., et~al.\ 1997, \apj\ 479, 222.

\bibitem[Mathur et~al.\ (2017)]{STORM07}
Mathur, S., Gupta, A., Page, K., et~al.\ 2017, \apj\ 846, 55.

\bibitem[McHardy et~al.\ (2018)]{MCH18}
McHardy, I.M., Connolly, S.D., Horne, K.D., Cackett, E.M., Gelbord, J., et~al.\ 2018, \mnras\ 480, 2881.

\bibitem[Mehdipour et~al.\ (2015)]{MEHD15}
Mehdipour, M., Kaastra, J.S, Kriss, G.A., Cappi, M., Petrucci, P.-O., et~al.\ 2015, A\&A 575, 22. 

\bibitem[Mehdipour et~al.\ (2016)]{MEHD16}
Mehdipour, M., Kaastra, J.S, Kriss, G.A., Cappi, M., Petrucci, P.-O., et~al.\ 2015, A\&A 588, 139. 

\bibitem[Moloney \& Shull (2014)]{MS14}
Moloney, J. and Shull, J.M.\ 2014, \apj\ 793, 100.

\bibitem[Mor \& Trakhtenbrot (2011)]{MT11}
Mor, R., \& Trakhtenbrot, B.\ 2011, \apj\ 737, L36.

\bibitem[Mor \& Netzer (2012)]{MN12}
Mor, R. \& Netzer, H.\ 2012, \mnras\ 420, 526.

\bibitem[Mor et~al.\ (2009)]{MNE09}
Mor, R., Netzer, H. and Elitzur, M.\ 2009, \apj\ 705, 298.

\bibitem[Morgan et~al.\ (2010)]{MOR10}
Morgan, C.W., Kochanek, C.S., Morgan, N.D., \& Falco, E.E.\ 2010, \apj\ 712, 1129.

\bibitem[Mosquera et~al.\ (2013)]{MOS13}
Mosquera, A.M., Kochanek, C.S., Chen, B., et al.\ 2013, \apj\ 769, 53.

\bibitem[Narayan, R.\ (1996)]{NR96}
Narayan, R.\ 1996, \apj\ 462, 136.

\bibitem[Netzer \&  Laor (1993)]{NL93}
Netzer, H., \& Laor, A.\ 1993, \apj\ 404, L51.

\bibitem[Netzer, H.\ (2015)]{NET15}
Netzer, H.\ 2015, ARA\&A, 53, 365.

\bibitem[Pancoast et~al.\ (2012)]{PAN12}
Pancoast, A., Brewer, B.J., Treu, T., et~al.\ 2012, \apj\ 754, 49

\bibitem[Pancoast et~al.\ (2014a)]{PAN14a}
Pancoast, A., Brewer, B.J., \& Treu, T.\ 2014a, \mnras\ 455, 3055.

\bibitem[Pancoast et~al.\ (2014b)]{PAN14b}
Pancoast, A., et al.\ 2014b, \mnras\ 445, 3073.

\bibitem[Pancoast et~al.\ (2018)]{PAN18}
Pancoast, A., Barth, A., Horne, K., Treu, T., Brewer, B.J., et~al.\ 2018, \apj\ 856, 108.

\bibitem[Pei et~al.\ (2017)]{STORM05}
Pei, L., Fausnaugh, M.M., Barth, A.J., Peterson, B.M., Bentz, M., et~al.\ 2017, \apj\ 837, 131, (Paper~{\sc v}).

\bibitem[Peterson et~al.\ (1991)]{P91}
Peterson, B.M., Balonek, T.J., Barker, E.S., et~al.\ 1991, \apj\ 368, 119.

\bibitem[Peterson et~al.\ (2013)]{PET13}
Peterson, B.M., Denney, K.D., De Rosa, G., Grier, C.J., Pogge, R. et~al.\ 2013, \apj\ 779, 109. 

\bibitem[Poindexter et~al.\ (2008)]{POIN08}
Poindexter, S., Morgan, N., \& Kochanek, C.S.\ 2008, \apj\ 673, 34

\bibitem[Ramolla et~al.\ (2018)]{Ramolla2018}
Ramolla, M., Haas, M., Westhues, C., et~al.\ 2018, A\&A 620, 137.

\bibitem[Rees et~al.\ (1989)]{RNF89}
Rees, M.J., Netzer, H., \& Ferland, G.J.\ 1989, \apj\ 347, 640.

\bibitem[Schlafly \& Finkbeiner (2011)]{SF11}
Schlafly, E.F., \& Finkbeiner, D.P.\ 2011, \apj\ 737, 103.

\bibitem[Shull et~al.\ (2012)]{SSD12}
Shull, J.M., Stevans, M., \& Danforth, C.W.\ 2012, \apj\ 752, 162.

\bibitem[Skielboe et~al.\ (2015)]{SK15}
Skielboe, A., Pancoast, A., Treu, T., Park, D., Barth. A.J., et~al.\ 2015, \mnras\ 454, 144.

\bibitem[Starkey et al.\ (2017)]{STORM06}
Starkey, D., Horne, K.D., Fausnaugh, M.M., Peterson, B.M., Bentz, M.C., et~al.\ 2017, \apj\ 835, 65, (Paper~{\sc vi}).

\bibitem[Suganuma et~al.\ (2006)]{SUG06}
Suganuma, M., Yoshii, Y., Kobayashi, Y., Minezaki, T., Enya, K. et~al.\ 2006, \apj\ 639, 46.

\bibitem[Walter et al.\ 1994]{Walter94}
Walter, R., Orr, A., Courvoisier, T.J.-L., Fink, H.H., Makino, F., Otani, C., \& Wamsteker, W.\ 1994, A\&A 285, 119.

\bibitem[Wamsteker et~al.\ (1990)]{WAM90}
Wamsteker, W., Rodriguez-Pascual, P., Wills, B.J., Netzer, H., Wills, D., et~al.\ 1990, \apj\ 354, 446.

\bibitem[White \& Peterson(1994)]{WP94}
White, R.J., and Peterson, B.M.\ 1994, PASP, 106, 879.

\end{thebibliography}




\appendix

\section{What lies beneath the Broad Emission Lines in AGN Spectra?}

The present paper explores the contributions of the continua emitted by the gas emitting the broad emission lines to the flux and lag spectra of a broad line AGN. We have shown that a substantial fraction of the flux within the continuum windows contains \textit{continuum} radiation from the BLR. However, in addition there are at least two other major emission regions associated with AGN which may contribute within a typical aperture centered on the AGN: the narrow emission line region and the dusty torus. The latter in particular is expected to have a substantial covering fraction of the central continuum source, and its grains both emit blackbody radiation and scatter a fraction ($\sim$20\%) of the incident near-UV-optical light (e.g., see Korista \& Ferland 1998). The gas, too, emits emission lines and thermal continuous radiation. Here we compute spectra from representative gas associated with these two regions lying outside the BLR, using constraints offered by the well-studied AGN, NGC~5548.

We utilize the two-component model of the NLR in NGC~5548 of Kraemer et~al.\ (1998).  The outer NLR has a hydrogen density of $\log \rm{n(H)} (cm^{-3}) = 4.5$, ionisation parameter $\log \rm{U(H)} = -2.5$, column density $\log \rm{N(H)} (cm^{-2}) = 21$, and distance from the central source of approximately 60 pc. The inner NLR has a hydrogen density of $\log \rm{n(H)} (cm^{-3}) = 7$, ionisation parameter $\log \rm{U(H)} = -1.5$, column density $\log \rm{N(H)} (cm^{-2}) = 21.5$, and distance from the central source of approximately 1 parsec. The two components have a combined covering fraction of the central source of ionising photons of $\approx 9\%$\footnote{More recently, Peterson et~al.\ (2013) and Kriss et~al.\ (2019) presented evidence that a substantial fraction of the gas emitting the narrow emission lines in NGC 5548 occurs within the gas participating in the outflow associated with the UV narrow absorption lines. This includes a component with a density n(H) $\approx 10^{5}$ cm$^{-3}$, lying 1-3 pc from the central source that is heavily obscured by gas (called the ``obscurer'') which might also be the UV broad absorber (Kaastra et~al.\ 2014; Arav et~al.\ 2015; Mehdipour et~al.\ 2016). This doesn't significantly change the findings here.}. Summed together, the predicted luminosities in narrow H$\beta$ $\lambda$4861 and [O III] $\lambda$5007 are as observed in NGC 5548 (Peterson et~al.\ 2013), with a flux ratio of about 0.11. Due to their low gas densities and covering fractions, together these two NLR components sum to only $\approx$1\% continuum contributions (but $\approx$2.5\% near the Balmer Jump) relative to the incident continuum light across the 1000-10,000\AA\/ spectral region\footnote{Unless otherwise noted, the quoted percentage contributions refer to the flux ratio F(diffuse)/F(incident).}. This source of continuum radiation forms a weak, very slowly varying background (e.g., see Peterson et~al.\ 2013). 

The toroidal obscuring region (TOR) represents a potentially more significant source of continuum radiation from outside the broad emission line region, particularly portions lying nearest to the central source (Mor \& Trakhtenbrot 2011; Mor \& Netzer 2012; Netzer 2015). Likely lying along the outer boundary of the BLR (Netzer \& Laor 1993; Baskin, Laor, \& Stern 2014; Baskin \& Laor 2018), graphite grains glowing at or near their sublimation temperatures (1700-1800~K, similar to tungsten filaments in incandescent light bulbs) \textit{and their finite UV-optical albedo} likely produce significant radiation in the near-UV and optical, in addition to in the near-IR. Unlike the for the NLR in NGC 5548, we do not have tight energetics constraints on this particular source of thermal and scattering radiation, and so here we adopt a conservative 25\% covering fraction of central source of radiation by the gas containing the hot grains (Mor, Netzer, \& Elitzur 2009; Mor \& Netzer 2012). 

In modeling this light source, we assume graphite grains with size distributions found in the Orion Nebula (Baldwin et~al.\ 1991), typically larger and so grayer than Milky Way ISM grains, appropriate in a harsh radiation environment. The hydrogen gas number density and column density are set to $10^8$ cm$^{-3}$ and $10^{25}$ cm$^{-2}$, respectively, while the illuminated face lies 126 light days from the central continuum source. At this distance the graphite grains lying near the illuminated face of the cloud are at or near their sublimation temperatures, for $\rm{L_{bol} \approx 10^{44.4}}$ ergs~s$^{-1}$ appropriate to NGC~5548. This model TOR's fractional contribution to the continuum ranges from 2-4\% in the near-UV (mainly due to grain scattering), to 3-5\% across the optical (4000\AA\/-7000\AA\/, mainly scattering with increasing thermal (gas and grain) contributions at longer wavelengths), rising in the near-IR to $\approx$28\% at 1 micron ($\mu\/m$), and becoming predominant at wavelengths beyond $\approx$1$\mu\/m$, mainly due to thermal radiation from the hot graphite grains. This source of continuum radiation, whose model spectrum peaks near $2\mu\/m$ in $\rm{\lambda\/F_{\lambda}}$ vs.\ $\rm{\lambda}$, is likely a significant contributor to the near-UV to near-IR continuum, and varies due to the reprocessing of flux variations in the central continuum on time scales of $\approx$ 50\% $\rm{R_{sublim}(graphite)/c}$. The near-IR delay has been measured at $\tau(2\mu\/m) \approx$ 60 days for NGC~5548; Koshida et~al.\ (2014), Suganuma et~al.\ (2006), while Landt et~al.\ (2019) determined a response-weighted mean $c<\tau> \approx$70 light days for dust in this AGN. 

\begin{figure}
\includegraphics[width=\columnwidth]{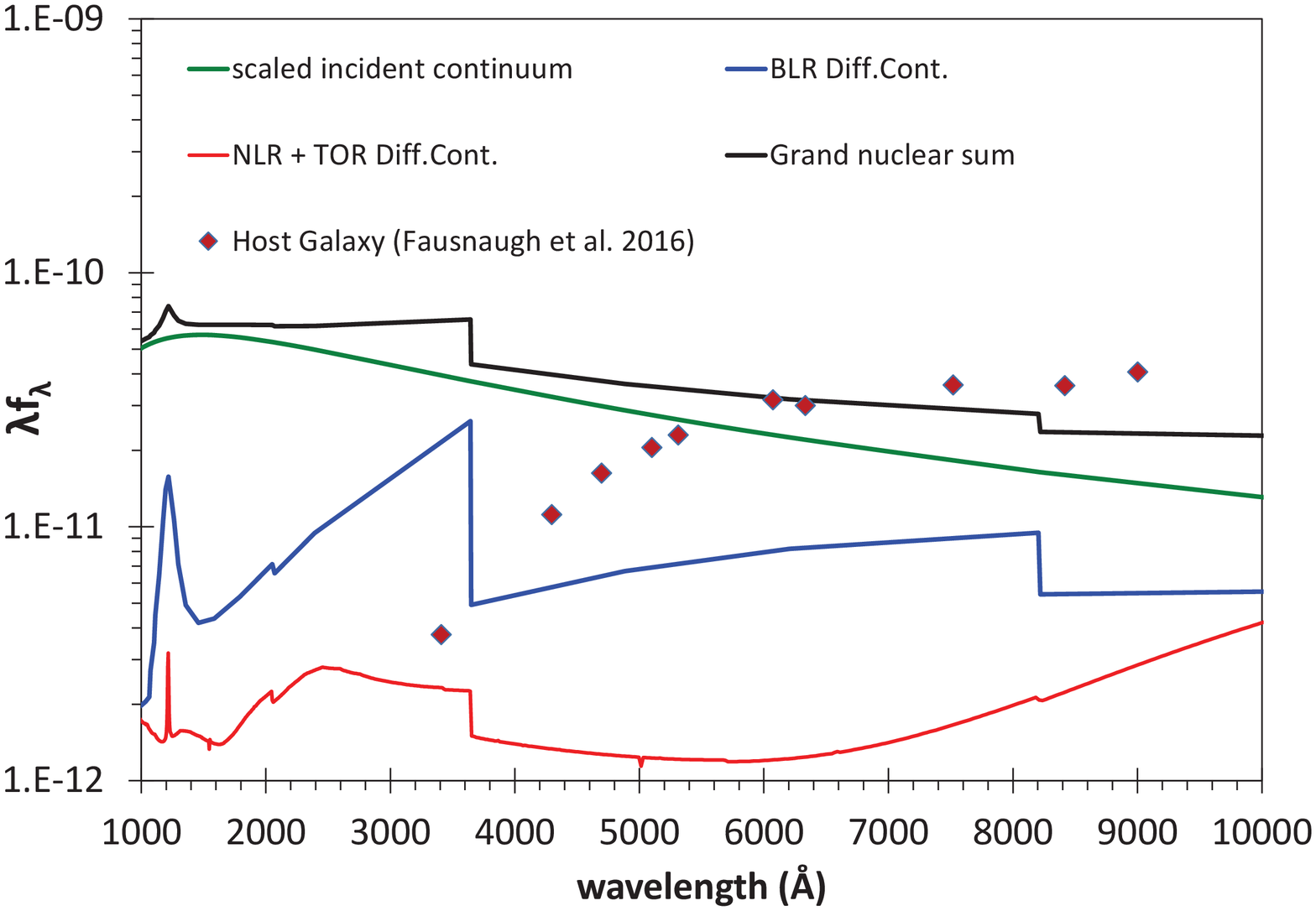}
    \caption{
    The computed diffuse continuum spectra (in units of erg~s$^{-1}$~cm$^{-2}$) of the inner and outer narrow line regions in NGC~5548 and dusty gas at a distance near the sublimation radius of graphite grains (red), along with the model of the BLR diffuse continuum from KG01 (blue) and a guess to the underlying accretion disk spectrum (Magdziarz et~al.\ 1998), and the sum (black). The emission lines have been omitted from the diffuse emission spectra for clarity.  See text for details.
    }
    \label{fig:appendix_figure1}
\end{figure}

\begin{figure}
\includegraphics[width=\columnwidth]{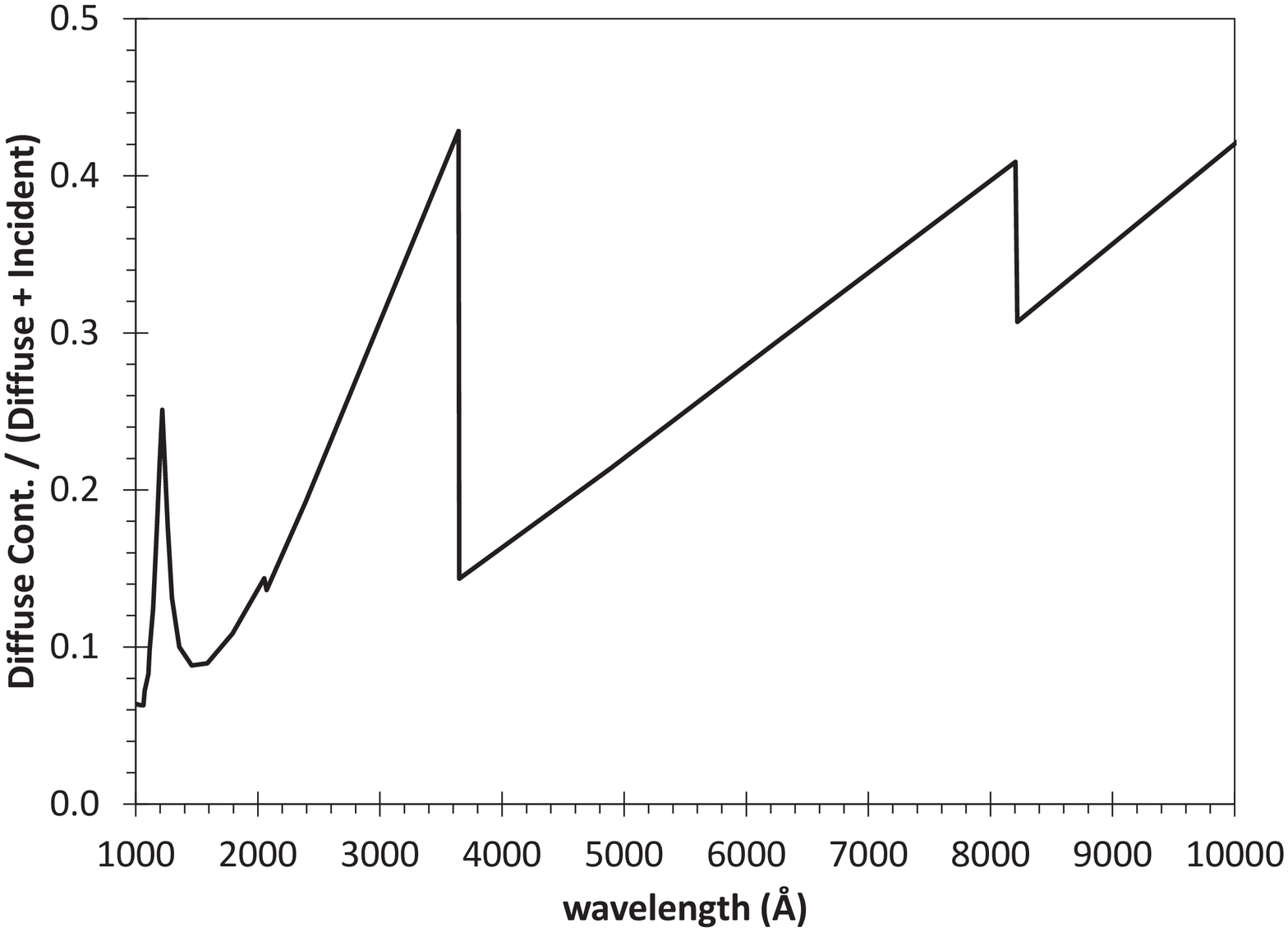}
    \caption{
The fractional contribution of the diffuse continua from the inner and outer NLR, TOR, and BLR relative to the total nuclear light. 
    }
    \label{fig:appendix_figure2}
\end{figure}

Figure~\ref{fig:appendix_figure1} shows the sum of the NLR and TOR spectral continuum components in red, that of the broad emission line region from the model of KG01 in blue, a reasonable estimate to the underlying nuclear continuum in green (that presented in Magdziarz et~al.\ 1998 for NGC~5548), and their sum in black for an AGN with a central luminosity of NGC~5548 ($\rm{L_{ion} \approx 10^{44.3}}$ ergs s$^{-1}$). The flux scale is an absolute one, with the estimate to the underlying accretion disk spectrum set such that its flux at $\lambda$1456 summed with the total diffuse continuum light at that wavelength (with about a 9\%, near minimum, contribution to the total nuclear light), together sum to the mean value in the spectrum of NGC~5548 for the 2014 monitoring campaign (Fausnaugh et~al.\ 2016; flux measurements corrected for Milky Way extinction). Also shown for comparison are photometric measurements of the host galaxy of NGC~5548 (Fausnaugh et~al.\ 2016) on the same flux scale. Finally, in Figure~\ref{fig:appendix_figure2}, we show the predicted wavelength-dependent fractional contributions of all aforementioned diffuse continua to the total nuclear light. The contributions are substantial --- from 10\% to 43\% across the rest-frame UV to near-IR spectrum, generally rising toward longer wavelengths. At wavelengths longer than $1\mu\/m$, the diffuse continuum from the BLR begins to decline, while as noted above that from the TOR continues to rise in $\rm{\lambda\/F_{\lambda}}$ out to $2\mu\/m$ or so.

It is not the intent, here, to present a detailed model fit to the spectrum of NGC 5548. Instead, we have presented representative UV to near-IR spectra of each of these sources of diffuse continuum radiation to raise awareness of their potentially important contributions to the UV to near-IR continuum flux and variability spectra of AGN.

\bsp	
\label{lastpage}
\end{document}